\documentclass[reprint,aps,prx,amsmath,amssymb,longbibliography,notitlepage,superscriptaddress]{revtex4-2}
\usepackage{amsmath,amssymb, bm,graphicx}
\usepackage{graphics}
\usepackage{float} 
\usepackage[stable]{footmisc}
\usepackage{soul}
\usepackage[colorlinks, linkcolor= blue, citecolor = blue, urlcolor=blue]{hyperref}
\usepackage{makecell}
\usepackage{physics}
\usepackage{notes2bib}
\def\be{\begin{equation}}
\def\ee{\end{equation}}
\def\bea{\begin{eqnarray}}
\def\eea{\end{eqnarray}}
\def\nn{\nonumber}
\begin{document}

\title{Nonlinear Landau fan diagram and aperiodic magnetic oscillations in three-dimensional systems}
%\title{Generalized Onsager's Relation: Nonlinear Landau fan diagram and aperiodic Quantum Oscillations in 3D systems}
\author{Sunit Das}
\email{sunitd@iitk.ac.in}
\affiliation{Department of Physics, Indian Institute of Technology Kanpur, Kanpur-208016, India}
\author{Suvankar Chakraverty} 
\email{suvankar.chakraverty@gmail.com}
\affiliation{Quantum Materials and Devices Unit, Institute of Nano Science and Technology, Sector-81, Punjab, 140306, India}
\author{Amit Agarwal}
\email{amitag@iitk.ac.in}
\affiliation{Department of Physics, Indian Institute of Technology Kanpur, Kanpur-208016, India}

\begin{abstract}
Quantum oscillations offer a powerful probe for the geometry and topology of the Fermi surface in metals. Onsager's semiclassical quantization relation governs these periodic oscillations in $1/B$, leading to a linear Landau fan diagram. However, higher-order magnetic susceptibility-induced corrections give rise to a generalized Onsager's relation, manifesting in experiments as a  nonlinear Landau fan diagram and aperiodic quantum oscillations. Here, we explore the generalized Onsager's relation to three-dimensional (3D) systems to capture the $B$-induced corrections in the free energy and the Fermi surface. We unravel the manifestation of these corrections in the nonlinear Landau fan diagrams and aperiodic quantum oscillations by deriving the $B$-dependent oscillation frequency and the generalized Lifshitz-Kosevich equation, respectively. Our theory explains the necessary conditions to observe these fascinating effects and predicts the magnetic field dependence of the cyclotron mass. As a concrete example, we elucidate these effects in a 3D spin-orbit coupled system and extract zero-field magnetic response functions from analytically obtained Landau levels. Our comprehensive study deepens and advances our understanding of aperiodic quantum oscillations. 
\end{abstract}

\maketitle

\section{Introduction}

Quantum magnetic oscillations, exemplified by the Shubnikov-de Hass (SdH) and de Hass-van Alphen (dHvA) effects, offer invaluable insights into the electronic structure, Fermi surface and topological properties of metals~\cite{Hass_30, shubnikov_30, Landau_30, lifshitz1956theory,lifshitz1958theory, dingle52, Sheonberg}. Driven by the population and depopulation of the quantized Landau levels (LLs) with varying magnetic field ($B$), these oscillations are generally periodic in  $1/B$~\cite{lifshitz1956theory, lifshitz1958theory, Glazman_prx18,glazman_fermiology23,Zhao_AP22}. The experimental probe of the variation of the oscillation peaks with $1/B$ produces a linear Landau fan diagram. This is well described by Onsager's semiclassical quantization relation~\cite{Onsager_52}, relating the LL index ($n$) to $1/B$. 
The slope of the Landau fan diagram determines the oscillation frequency that maps the Fermi surface of the metal~\cite{lifshitz1956theory,lifshitz1958theory}, and the intercept to the $n$ axis determines its topological properties~\cite{Mikitik_99,Novoselov2005, Zhang2005, Shen_science09,Fuchs_epj10, Tokura_13, wright_prb13, Glazman_prx18, genfu_prx15, gao_PNAS17, Mandar_19, Zhao_AP22,glazman_fermiology23,Binghai_23quantum}. Another important experimental feature is  the wave-form of the quantum oscillations, which is well described by the Lifshitz-Kosevich (LK) equation 
%, which also enables the estimation of the effective cyclotron mass and  mobility
~\cite{lifshitz1956theory,lifshitz1958theory,dingle52, Sheonberg}. 

\begin{figure}[t!]
    \centering
    \includegraphics[width=.99\linewidth]{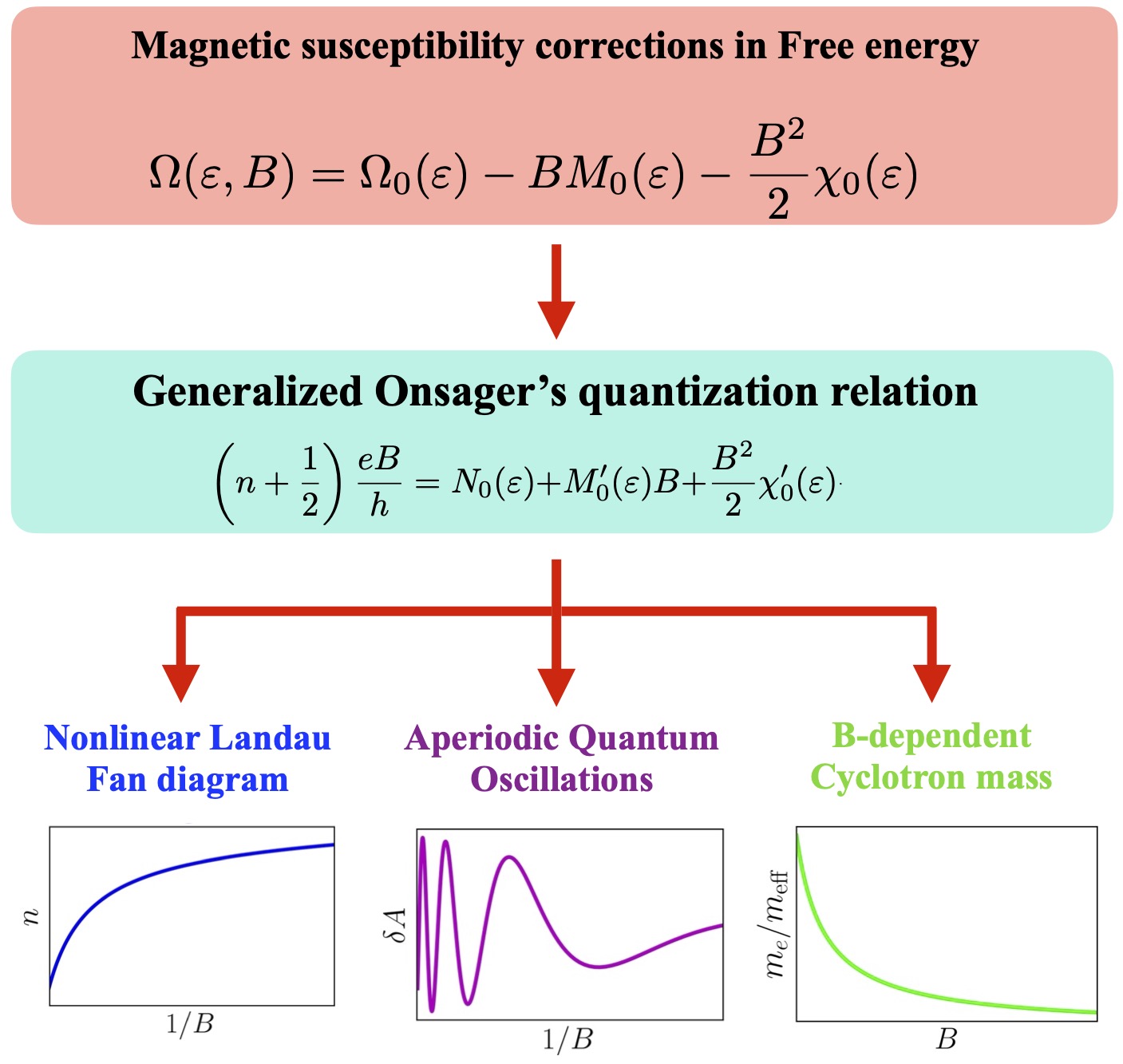}
    \caption{The schematic highlights the origin of the generalized Onsager's quantization relation and its consequences. Physically, the generalization results from the magnetic field-induced modification of the Fermi surface and the magnetic susceptibility corrections in the Free energy. These corrections lead to three dominant effects in experiments: i) A nonlinear Landau fan diagram, where the relation between the Landau level index ($n$) and $1/B$ (inverse magnetic field strength) deviates from the conventional straight line. 
    ii) Aperiodic magnetic oscillations in physical quantities (denoted as $A$) with $1/B$, where the frequency of oscillations starts varying with the magnetic field. iii) The effective cyclotron mass ($m_{\rm eff}$) becomes magnetic field-dependent. 
    \label{fig_1}}
\end{figure}

However, the observation of peculiar deviations from periodicity in SdH oscillations and nonlinear Landau fan diagram in SrTiO$_3$ and KTaO$_3$-based electronic interfaces~\cite{Yang_apl16,Fete_njp14,Nini_prl16,Zeitler_prr21, Ong_science10,ariando_ami22, Rubi_npj20,rubi_2023unconventional} and in the surface states of topological insulators~\cite{ando_prb11,ando_prb10_Bi2Te2Se,Ong_science10,wright_prb13}, challenges the traditional understanding. A potential origin of the nonlinear Landau fan diagram is the $B$-induced corrections to the band energy and the Bloch wave function, which also reflects in a change in the free energy and the Fermi surface~\cite{Gao_prl14,Gao_prb15,Gao_FP19}. Despite recent efforts to understand this in 2D materials~\cite{gao_PNAS17, Fuch_scipost18}, a comprehensive understanding of observed aperiodic oscillations in 3D systems is still lacking~\cite {Roth66}. For instance, Onsager's relation was recently generalized for 2D systems to include the zero-field magnetic response function~\cite{gao_PNAS17, Fuch_scipost18}. %Consequently, due to the presence of higher-order $B$ terms, the generalized Onsager's relation creates a nonlinear Landau fan diagram~\cite{Gao_prl14, Fuch_scipost18}. 
Nevertheless, the experimental consequences of this generalization have not been fully explored. 

To remedy this, we generalize Onsager's semiclassical quantization for 3D systems to include $B$-induced corrections to the Fermi surface. Here, we demonstrate its impact on nonlinear Landau fan diagrams and aperiodic quantum oscillations. We systematically study the experimental consequences of magnetic field-induced Fermi surface modification [see Fig.~\ref{fig_1}] and address the following key aspects. i) What are the conditions for observing a nonlinear Landau fan diagram? ii) How do we generalize the Lifshitz-Kosevich equation to capture aperiodic quantum oscillations? iii) What is the magnetic field dependence of the effective cyclotron mass arising from $B$-induced corrections in the Fermi surface area? Additionally, we explicitly demonstrate the consequences of $B$-induced Fermi surface modifications in a 3D spin-orbit coupled (SOC) system via exact calculations.  
Our work deepens our understanding of quantum magnetic oscillations and gives us access to a new probe for measuring more nuanced features of the Fermi surface and magnetic response functions~\cite{Slot_science23}. 

%%%%%%%%%%%%%%% organization of the paper %%%%%%%%%%%%%%%

The rest of our paper is organized as follows. We discuss the generalization of Onsager's relation for 2D and 3D systems in Sec.~\ref{Sec_gen_Onsager}. In Sec.~\ref{Sec_expt_observables} and~\ref{Sec_observables_QM}, we explore the experimentally observable signatures of aperiodic oscillation in the semiclassical and quantum mechanical framework, respectively. We demonstrate the nonlinear fan diagram, aperiodic oscillation, and $B$-dependent cyclotron mass for a 3D SOC system in Sec.~\ref{Sec_3DSOC}. 
Finally, we discuss some subtleties and recent experiments on aperiodic oscillations in Sec.~\ref{Sec_discussion} and follow it up with a summary of our results in Sec.~\ref{Sec_conclusion}.

\begin{figure*}
    \centering
    \includegraphics[width=\linewidth]{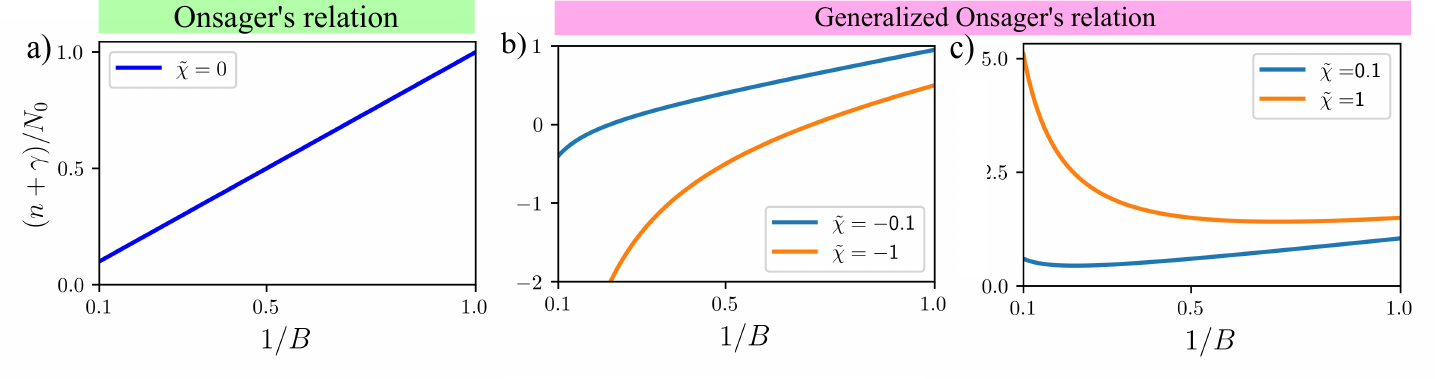}
    \caption{The variation of the Landau level index $\frac{(n+\gamma)}{N_0}$ with $1/B$, for a) $\tilde \chi=0$ supporting a linear Landau fan diagram, b) two different values of negative $\tilde{\chi}$ and c) positive $\tilde \chi$, respectively, showing a nonlinear Landau fan diagram.  The parameter $\tilde \chi$ captures the $B$-induced modification of the Fermi surface, giving rise to the nonlinearity in the Landau fan diagram.  %The plots reveal that increasing $\tilde \chi$ leads to a more nonlinear Landau fan diagram, whereas $\tilde \chi=0$ plot corresponds to the conventional Onsager relation, producing a linear Landau fan diagram. 
    }
    \label{fig0}
\end{figure*}

\section{Generalized Onsager's semiclassical quantization relation} 
\label{Sec_gen_Onsager}
This section first discusses the conventional Onsager's semiclassical quantization relation. Then, the generalization of the quantization relation, incorporating the higher-order magnetic response functions for 2D and 3D systems, has been discussed. 

\subsection{Semiclassical Onsager's relation and the linear Landau fan diagram}
Semiclassically, the quantization of the cyclotron orbits is represented by the conventional Onsager's relation specified by, 
\be \label{onsager}
\frac{S_0(\varepsilon)}{4\pi^2} \equiv N_0(\varepsilon) = \left( n+\frac{1}{2} \right) \frac{e B}{h}~.
\ee 
Here, $N_0$ is the total number of charge carriers in the system, or alternately, the zero-field integrated density of states (IDOS) up to energy $\varepsilon$. $S_0$ is the area of the cyclotron orbit. Equation~\eqref{onsager} implies that the energy level index varies linearly with the inverse of the magnetic field, i.e.,  $n \propto 1/B$. This produces a linear Landau fan diagram and manifests as magnetic oscillations in physical quantities being periodic in $1/B$. The frequency of magnetic oscillation is proportional to the $N_0(\varepsilon_F)$, where $\varepsilon_F$ is the Fermi energy. 

This paradigm has been immensely useful for understanding the Fermi surface of metallic systems by measuring either SdH oscillations in magneto-transport or the dHvA oscillations in magnetization. However, a core assumption in Eq.~\eqref{onsager} is that the Fermi surface, or the electronic states, either does not change under the application of a magnetic field or the change is small enough to manifest in experiments. This motivates a natural question: what happens when this assumption does not hold?

\subsection{Generalized Onsager's relation and nonlinear Landau fan diagram}

The semiclassical Onsager's relation neglects the magnetic field-induced corrections to the electronic states or the IDOS, $N_0(\varepsilon)$. However, the semiclassical IDOS and the energy may get modified by magnetic field-dependent corrections in a magnetic field~\cite{Gao_prl14,Gao_prb15,Fuch_scipost18}. 
An intuitive way to understand this is that the IDOS can be obtained from the semiclassical free energy $\Omega_{\rm semi}$, using the relation $N_0=-\partial_{\varepsilon} \Omega_{\rm semi}$. In the presence of a magnetic field, the non-oscillating part of the free energy, i.e., $\Omega_{\rm semi}$, becomes magnetic field-dependent. The magnetic corrections in $\Omega_{\rm semi}$ are specified by the zero-field magnetic response functions multiplied by the powers of the magnetic field. In general, the semiclassical free energy admits a series expansion in powers of $B$, and we have~\cite{Fuch_scipost18}
\be 
\Omega_{\rm semi}(\varepsilon, B) =  \Omega_0(\varepsilon) - B M_0(\varepsilon) -\frac{B^2}{2} \chi_0(\varepsilon) - \sum_{p \geq 3} \frac{B^p}{p!} R_p(\varepsilon) .
\ee 
As a result, magnetic field dependent $N_0(\varepsilon)$ is given by,  
\be \label{N_B}
N(\varepsilon,B) = N_0(\varepsilon) +  B M'_0(\varepsilon) + \frac{B^2}{2} \chi'_0(\varepsilon) + \sum_{p \geq 3} \frac{B^p}{p!} R'_p(\varepsilon).
\ee 
Here, $ M'_0 (\varepsilon)$ is the energy derivative of the zero-field spontaneous magnetization and contains the contribution from the total magnetic moment (orbital and spin) and the Berry phase. The $\chi'_0 (\varepsilon)$ is the energy derivative of the zero-field magnetic susceptibility, and $R_p(\varepsilon)$ represents higher-order magnetic response functions of the system calculated at the zero temperature. When the above modification of $N_0(\varepsilon) \to N(\varepsilon, B)$ in Eq.~\eqref{onsager}, the conventional semiclassical quantization relation yields the generalized Onsager's relation or the `Roth-Gao-Niu' quantization condition. We note that historically, Roth generalized Onsager's relation to the second order in the magnetic field~\cite{Roth66}. Recently, Gao and Niu obtained the modified Onsager's relation by systematically including all the higher-order corrections in magnetic field~\cite{gao_PNAS17}. Using Eq.~\eqref{N_B} in Eq.~\eqref{onsager}, the generalized Onsager's quantization relation can be expressed as~\cite{Roth66, gao_PNAS17, Fuch_scipost18} 
%
%\bea \label{modified_onsager}
%\left(n+\frac{1}{2} + \frac{h}{e} M'_0(\varepsilon) \right)\frac{eB}{h} 
% &&=N_0(\varepsilon) + \frac{B^2}{2} \chi'_0(\varepsilon) \nn \\
% && +  \sum_{p \geq 3} \frac{B^p}{p !} R'_p(\varepsilon) .
%\eea 
\be \label{modified_onsager}
\left(n+\frac{1}{2} \right)\frac{eB}{h}=N_0(\varepsilon) +  M'_0(\varepsilon) B + \frac{B^2}{2} \chi'_0(\varepsilon) +  \sum_{p \geq 3} \frac{B^p(\varepsilon)}{p !} R'_p .
\ee 
 From the above equation, we can infer that the $M'_0(\varepsilon)$ term modifies the phase of the magnetic oscillation while retaining the relation $n \propto N_0(\varepsilon)/B$, i.e., linear $n$ vs. $1/B$ relation. In Eq.~\eqref{modified_onsager},   The  $\chi'_0 (\varepsilon)$ term is the lowest order correction in conventional Onsager's relation that introduces a nonlinear relationship between $n$ and $1/B$. A nonlinear $n$ vs. $1/B$ relation directly indicates the magnetic oscillation to become nonlinear or aperiodic in $1/B$. We discuss this in more detail in  Sec.~\ref{Sec_modified_LK}. %Interestingly, the modified Onsager's relation in Eq.~\eqref{modified_onsager} can also be expressed in the form of Eq.~\eqref{onsager} as $\left[n + \gamma(\varepsilon, B) \right]\frac{eB}{h} = N_0(\varepsilon)$~\cite{Roth66, Fuch_scipost18}, with the energy and magnetic field-dependent phase-shift is given by the expression
%
% \be \label{gamma}
% \gamma(\varepsilon,B)=\frac{1}{2} -  \frac{h}{e} M'_0(\varepsilon) - \frac{h B}{2e} \chi'_0(\varepsilon) -  \sum_{p \geq 3} \frac{h B^{p-1}}{e p !} R'_p(\varepsilon)~.
% \ee
% %
% Here, the $1/2$ is associated with the Maslov phase, which is $\pi$ for simply closed orbits~\cite{Fuchs2010}.  

% The modification of the Onsager relation can be explained using the semiclassical free energy of the system.   

\subsection{Generalized Onsager's relation for three dimensional systems}

 Even though the generalized Onsager's relation in 2D systems has been discussed at length in Refs.~\cite{gao_PNAS17, Fuch_scipost18}, the generalization for 3D systems is relatively less explored~\cite{Roth66}. 
In contrast to 2D systems, for a three-dimensional (3D) system, the energy levels along the magnetic field direction (say, along $\hat{z}$) have finite components that are dispersive and not quantized. Consequently, the Onsager's relation and the cyclotron orbit area (in the $k_x$-$k_y$ plane) become $k_z$ dependent. Considering closed cyclotron orbits,  
Onsager's quantization condition describes the relation between the cyclotron orbit area $S(\varepsilon,k_z)$ at each $k_z$ with the LL index $n$. This is given by
\be \label{onsagar_3d}
\frac{S(\varepsilon,k_z)}{4\pi^2}\equiv N_0(\varepsilon,k_z) =\left( n+\frac{1}{2} \right) \frac{eB}{h}~.
\ee  
Here, the $S(\varepsilon,k_z)$ is the cross-sectional area of the orbit in the ${\bm k}$-space whose boundary is defined by the intersection of constant energy surface with  $k_z={\rm constant}$ plane. Similar to the two-dimensional case, generalizing Onsager's relation for the three-dimensional case involves incorporating zero-field magnetic response functions in $N_0(\varepsilon,k_z)$. 
% The $\gamma(\varepsilon,k_z,B)$ is the $k_z$-dependent phase, which becomes $B$-dependent if we include the $k_z$-resolved zero-field magnetic response functions. 
Consequently, the magnetic field-dependent $k_z$ resolved IDOS is given by
\bea 
N(\varepsilon,k_z,B)= &&  N_0(\varepsilon,k_z) + B M'_0(\varepsilon,k_z) +\frac{ B^2}{2} \chi'_0(\varepsilon,k_z) \nn \\
&& +  \sum_{p \geq 3} \frac{B^{p}}{ p !} R'_p(\varepsilon,k_z)~.
\eea
%
%Note that historically, Roth first obtained the expression of $\gamma(\varepsilon,k_z, B)$ to first order in $B$, which includes the susceptibility term~\cite{Roth66}. Similar to the generalized Onsager's relation for 2D in Eq.~\eqref{modified_onsager}, we can express the modified Onsager's relation (or Roth-Gao-Niu quantization criteria) for 3D system as
Using the above expression in Eq.~\eqref{onsagar_3d}, we obtain the modified Onsager's relation as follows:
\bea \label{Modified_onsager_3D}
\left( n+\frac{1}{2} \right)\frac{eB}{h} =&& N_0(\varepsilon,k_z)   + B M'_0(\varepsilon,k_z) + \frac{B^2}{2} \chi'_0(\varepsilon,k_z) \nn \\
&& +  \sum_{p \geq 3} \frac{B^{p}}{p !} R'_p(\varepsilon,k_z)~.
\eea
%
% Here, we have used $S(\varepsilon,k_z)\equiv N_0(\varepsilon,k_z)$, the $k_z$-resolved IDOS for the system without a magnetic field. 
In contrast to the 2D case, the magnetic response functions in the above equation are $k_z$-dependent. %Hence, these are not the physical response functions; rather, 
%They represent the corresponding values at the plane of cyclotron motion. 
We can evaluate the physical response function by integrating with respect to $k_z$ to add the contribution from the whole Fermi volume. The susceptibility derivative  $\chi'_0 (\varepsilon,k_z)$ is the lowest order term that gives rise to aperiodic magnetic oscillations and nonlinear Landau fan diagrams in 3D systems. Typically, it is also the most dominant contribution. 
%However, in contrast with the two-dimensional case, the modified Onsager's relation provides the $k_z$-dependent zero-field response function for a three-dimensional system, as evident from the above equation. 

The interpretation of the Landau fan diagram in 3D systems is a bit subtle as Onsager's relation is $k_z$ dependent. %In 2D systems, Eq.~\eqref{onsager} shows that the slope of the linear dependence of $n$ on $1/B$ is dictated by $N_0(\varepsilon)$. However, in 3D systems, it is dictated by $N_0(\varepsilon,k_z)$. 
In contrast to the complete level occupancy or vacancy for 2D systems, the filling of each energy level changes smoothly with the changes in the magnetic field strength for 3D systems. However, at specific values of $k_z$ (say, $k_z=k_{zm}$), the variation of energy with respect to the $k_z$ becomes extremal i.e., $\partial_{k_z}\varepsilon_n(k_z)=0$. This extremal energy condition is equivalent to the extremization of the cyclotron orbit area~\cite{glazman_fermiology23}. For these extremal cases, the oscillating density of states (DOS) has maxima or minima. Consequently, physical observables sensitive to DOS also show peaks or troughs in low-temperature experiments. The extremal Fermi surface area corresponding to the turning points in the energy spectrum along $k_z$ contributes predominantly to the magnetic oscillations~\cite{lifshitz1956theory,lifshitz1958theory, Sheonberg, Roth66}. 
Consequently, the experimentally obtained Landau fan diagram captures the extremal orbit area via the coefficient of the $1/B$ term in $n$ vs. $1/B$ plot. 

We note that for both the 2D and 3D systems, the generalized Onsager's relation is valid for each carrier species or each electron and hole pocket separately. Hence, to obtain total physically relevant response functions, we need to sum up all the carrier species in the system, such as different bands, valleys, spin, etc. %Additionally, for a three-dimensional system, we must perform a $k_z$ integration of the $k_z$ dependent response functions to include the contribution from the whole Fermi surface. 

%Finally, we can model aperiodic oscillations in a system based on Eq.~\eqref{modified_onsager} (for the 2D system) or~\eqref{Modified_onsager_3D} (for the 3D system) by explicitly calculating all the zero-field magnetic response functions. In an independent and complementary approach, we can calculate the LLs (LLs) energies of a given system and invert the obtained energies to express the LL index $n$ as a function of energy and the magnetic field (also, $k_z$ for a 3D system). By expanding the obtained expression in powers of $B$, we can extract all the zero-field magnetic response functions. This offers an independent way to calculate all the zero-field magnetic response functions by comparing the expansion of the LL index in powers of $B$ with Eq.~\eqref{modified_onsager} or~\eqref{Modified_onsager_3D}~\cite{gao_PNAS17, Fuch_scipost18}. This approach was recently used in Refs.~\cite{Tisserond_IOP17,rubi_2023unconventional} to explain experimental observation of aperiodic oscillation in different two-dimensional systems. In this paper, we adopt the latter approach to explain the aperiodic magnetic oscillation in 3D SOC systems for the first time to the best of our knowledge. We also show the explicit calculations for the 2D SOC system in Appendix~\ref{2D_rashba_sec}.

\section{Observable signatures of generalized Onsager's relation}\label{Sec_expt_observables}
%of aperiodic magnetic oscillation }
\label{sec3}

Having established the generalized Onsager's quantization relation, we now discuss the consequences of generalized Onsager's quantization relation, focusing on the experimentally observable effects. We first discuss the conditions for the appearance of nonlinearity or aperiodicity in the Landau fan diagram. We follow this by discussing the impact of the magnetic susceptibility corrections on the modified waveform of oscillating physical quantities (generalized Lifshitz-Kosevich formula) and the magnetic field-dependent effective mass extracted from magnetoresistance or thermodynamic magnetization measurements. 

%However, before that, we mention the role of electron density in the system for observing aperiodic oscillation. %We discuss this in view of having a nonlinear Landau fan diagram. While we focus on the two-dimensional system below, a similar argument works for three-dimensional systems.
%Following the modified Onsager's relation~\eqref{modified_onsager}, the lowest order correction introducing aperiodicity in magnetic oscillation is the energy derivative of the susceptibility at the Fermi energy. 
%\textcolor{blue}{SD: is the above claim right? need to discuss with experimentalist}
%The conventional Onsager's quantization relation produces a linear relationship between the LL index $n$ and the inverse of the applied magnetic field $B$, known as the Landau fan diagram. However, the Landau fan diagram becomes nonlinear owing to the zero-field magnetic response functions in the modified Onsager's relation.

\subsection{Nonlinear Landau fan diagram} \label{Sec_nl_Fan}

First, let us retain up to quadratic $B$ terms in Eq.~\eqref{modified_onsager}. We have,  
$(n+1/2)~\frac{eB}{h} = N_0(\varepsilon)+M'_0(\varepsilon) B + \frac{B^2}{2} \chi'_0(\varepsilon)$. We rewrite this as 
\be \label{Onsager_upto_chi}
\frac{e/h}{N_0(\varepsilon)} \left[ n+\gamma(\varepsilon)  \right] = \left[\frac{1}{B} + \frac{B}{2} \frac{\chi'_0(\varepsilon)}{N_0(\varepsilon)} \right]~.
\ee
Here, we have defined, $\gamma(\varepsilon)=\frac{1}{2}-\frac{h}{e} M'_0(\varepsilon)$. From Eq.~\eqref{Onsager_upto_chi}, it is evident that the predominant factor dictating the aperiodicity of magnetic oscillations is $\tilde{\chi} = \chi'_0/N_0$, at the Fermi energy. Systems in which we have $\tilde{\chi} B^2$ to be of the order of one are likely to show aperiodic quantum oscillations even with closed orbits and no magnetic breakdown~\cite{moon2023nonlinear}. 
This can occur if $\chi'$ diverges at the chemical potential or if $N_0$, the carrier density in the system is relatively small. To demonstrate this, we show schematic plot the $(n+\gamma)/N_0$ vs. $1/B$ (putting $e=\hbar=1$) for different values of $\tilde{\chi}=\chi'_0/N_0$ in Fig.~\ref{fig0}. 

Our analysis implies the following: i) with increase in $\tilde \chi$, the nonlinearity in $(n+\gamma)/N_0$ vs. $1/B$ curve increases; (ii) for a particular value of $\tilde \chi$, the aperiodicity becomes pronounced for higher magnetic field values; (iii) the curvature of the nonlinear Landau fan diagram is determined by the sign of $\tilde \chi$, i.e., $\chi'_0$. Specifically, the negative $\chi'_0$ produces a concave-down Landau fan diagram, whereas the positive $\chi'_0$ produces a concave-up fan diagram. Furthermore, since $N_0$ captures the extremal Fermi surface area, the $B$-induced correction (specified by $\tilde \chi B^2/2$) captures the magnetic field-induced expansion or contraction of the Fermi surface. Thus, if the Fermi pocket itself is small, then the magnetic field-induced corrections to the Fermi pocket will likely have more impact on the observable properties.  

%An intuitive way to think about these corrections showing up in experiments

%Hence, the $N_0$, being proportional to the electron density, plays a crucial role in determining the degree of aperiodicity of oscillation in a system, along with the $\chi'_0$. For high-carrier density, the aperiodicity factor $\tilde \chi=\frac{\chi'_0 (\varepsilon_F)}{ 2 N_0(\varepsilon_F)}$ becomes small, which may result in periodic oscillation in the system. 
 
% We discuss these with specific models: 2D and 3D SOC systems in Appendix~\ref{2D_rashba_sec} and Sec.~\ref{sec5}, respectively.

% To substantiate our above analysis with the experimental observations, we highlight the Shubnikov de-Hass oscillation observed in a KTaO$_3$-based conducting interface (Euo-KTaO$_3$). 

The Landau fan diagram is typically extracted from the peaks and troughs of the magneto-resistance or magnetization measurements. This prompts the question: what is the impact of the higher-order magnetic susceptibility terms on the waveform of these measurements? We address this in the following subsection by focusing on the waveform of the oscillating part of the physical quantities. 

% \textcolor{red}{I am unsure of the above explanation. Theoretically, our explanation seems okay, but experimentally, things are not so consistent.}

% Theoretically, we can obtain $n$ vs. $1/B$ plot in a system based on Eq.~\eqref{modified_onsager} by explicitly calculating all the zero-field magnetic response functions. In an independent and complementary approach, we can calculate a system's exact Landau level (LL) energies and invert the obtained energies to express the LL index $n$ as a function of energy and the magnetic field. By expanding the expression obtained in powers of $B$, we can extract all the zero-field magnetic response functions. This approach is advantageous when we have the analytically calculated LLs. This method was recently used to explain the aperiodic oscillation in $\alpha$-I$_3$ organic metal~\cite{Tisserond_IOP17}, complex oxide interfaces~\cite{rubi_2023unconventional}\ref{}. This paper adopts the latter approach to explain aperiodic oscillation in 2D and 3D SOC systems.

%\textcolor{blue}{SD: can we add a few lines about determining phase from the fan diagram in case of aperiodic oscillation?
%The $nB$ vs. $B^2$ plot of PNAS is not general; will think about it...}

\subsection{Generalized Lifshitz-Kosevich formula}\label{Sec_modified_LK}
In the presence of a strong magnetic field, quantized Landau levels are formed. 
%Consequently, physical quantities oscillate with varying magnetic fields. This occurs due to the fact that as we
On increasing (decreasing) the magnetic field, the LLs get emptied (filled) with the carriers. These changes in the LL fillings manifest in terms of oscillation in physical quantities. In the semiclassical limit, where many distinguishable quantized LLs are occupied, the expression of the oscillating part of the physical quantities was derived by Lifshitz and Kosevich, based on the conventional Onsager's quantization relation~\cite{lifshitz1956theory,lifshitz1958theory}. In this limit, the $k$-space orbit radius is much larger than the magnetic length scale $l_B=\sqrt{\hbar/eB}$. Consequently, using Poisson's summation formula [Eq.~\eqref{Poisson_sum}], the contribution of all occupied LL can be expressed as a sum over the Fourier series harmonics. Using this, we can express the oscillating part of the longitudinal magnetoresistivity or the magnetization for a 3D system as~\cite{lifshitz1956theory,lifshitz1958theory, Binghai_23quantum}
\bea \label{osc_amp_onsager}
&& \delta A  \propto  \sum_p \int dk_z A_{p}(k_z) \cos\left(2\pi p n + \varphi_A \right), \\
&& \propto  \sum_p \int dk_z A_{p}(k_z) \cos\left[2\pi p\left(\frac{h S(\varepsilon,k_z)}{eB}  -\frac{1}{2} \right) +\varphi_A \right]. \nn
\eea 
{Here, we mention that for simplicity, we have ignored the damping effects in the oscillation since we focus on the modification in the wave-form of oscillating quantities due to generalization in Onsager's relation.} In the second line, we used the LL index $n$ expression as specified by Eq.~\eqref{onsagar_3d}. In Eq.~\eqref{osc_amp_onsager}, $A$ is the oscillating physical quantity, with $\delta A$ representing its oscillating waveform, and $\varphi_A$ is the phase associated with the measured quantity. $A_p(k_z)$ is the oscillation amplitude for $p$th harmonic, which generally depends on $k_z$ for a three-dimensional system. In practice, only the fundamental harmonics ($p=1$) contribute predominantly to the oscillation, owing to the temperature and disorder-induced damping, which increases for higher-harmonics~\cite{Sheonberg,dingle52}. %The $k_z$ integration is usually performed around the extremal $k_z$ since $S(\varepsilon,k_z)$, for which the Fermi surface orbit area becomes maximum or minimum.
Nonetheless, the oscillation in physical quantities is periodic in $1/B$ for all the harmonics. In experiments, the oscillation period maps the extremal zero-field Fermi surface orbit area $S_m(\varepsilon)$. This is because although the $S(\varepsilon,k_z)$ is $k_z$ dependent, the maximum contribution in the integral comes from the extremal cyclotron orbits on the Fermi surface. See Refs.~\cite{lifshitz1956theory,lifshitz1958theory,Sheonberg,Roth66} for more details. 

As discussed in Sec.~\ref{Sec_gen_Onsager}, incorporating the magnetic response functions in the free energy leads to generalized Onsager's relation, and the LL index $n$ contains higher-order corrections in $B$. Consequently, the conventional LK formula also gets modified to account for the higher-order magnetic response functions in the oscillating part of physical quantities. Considering the modified Onsager's quantization relation up to the second order in $B$, we derive the oscillating part of the LK equation. Using Eq.~\eqref{Modified_onsager_3D} in Eq.~\eqref{osc_amp_onsager}, we obtain
\begin{widetext}
\begin{subequations}
\bea 
\delta A \propto && \sum_p \int dk_z A_{p}(k_z) \cos \left[2\pi p \left \{\frac{h }{eB} \left( N_0(\varepsilon,k_z) + B M'_0(\varepsilon,k_z) + \frac{B^2}{2} \chi'_0(\varepsilon,k_z) \right) -\frac{1}{2} \right \}+\varphi_A \right], \\
\label{modified_LK}
\propto && \sum_p \int dk_z A_{p}(k_z) \cos \left[2\pi p \left \{\frac{h }{e B} \left( N_0(\varepsilon,k_z) + \frac{B^2 }{2} \chi'_0(\varepsilon,k_z)  \right) + \frac{h}{e} M'_0(\varepsilon,k_z) -\frac{1}{2} \right \}+\varphi_A \right] .
\eea 
\end{subequations}
\end{widetext}
In the last equation, we have rearranged the terms to highlight that the oscillation frequency has an additional $B$-dependent term associated with the $\chi'_0$ (and other higher-order response functions, which we have not explicitly written here). 
This makes the oscillation in the measured magneto-resistance or magnetization to be nonlinear and aperiodic in $1/B$. The zero-field spontaneous magnetization term $ M'_0$ modifies only the oscillation phase in the measured quantity. 

The magnetic field-dependent oscillation frequency $F(B)$ can be derived by taking the derivative with $1/B$ of the $B$-dependent terms in the argument of the cosine function~\cite{rubi_2023unconventional}. We find that $F(B) \equiv N_{0m} - B^2 \chi'_{0m}/2$, where $N_{0m}$ and $\chi'_{0m}$ represent the values of the corresponding quantities, for extremal points on the Fermi surface, where $\partial_{k_z} \varepsilon_n(k_z) = 0$. Here, $N_{0m}$ represents conventional oscillation frequency, which captures the zero-field extremal Fermi surface orbit area. The term $- B^2 \chi'_{0m}/2$ is responsible for the magnetic field-dependent oscillation frequency, and it also estimates the magnetic field-induced change in the orbit area. 

In any system showing a nonlinear Landau fan diagram due to magnetic susceptibility-induced correction, the experimentally measured waveform of aperiodically oscillating physical quantities should be fitted with Eq.~\eqref{modified_LK}. More importantly, the fitting of the observed LL fan diagram with Eq.~\eqref{Modified_onsager_3D} and the fitting of the observed physical quantities should yield the same values of parameters for $N_{0m}$, $M_{0m}'$, and $\chi_{0m}'$, within experimental tolerance. 
 %considering the additional linear $B$ term. The recent study of aperiodic Shubnikov de Hass oscillation on KTaO$_3$-based electronic interface corroborates our proposal (personal communications). 

\subsection{Magnetic field dependent effective cyclotron mass} \label{effective_mass_B}

In addition to the nonlinear Landau fan diagram and the aperiodic oscillation in the magneto-resistance or magnetization, the magnetic field-induced changes in the Fermi surface also make the measured cyclotron mass of the system $B$ dependent. To understand this, we note that the cyclotron mass for a charge carrier is defined as $m_{\rm eff}=\frac{\hbar^2}{2 \pi}\left[ \frac{\partial S(\varepsilon)}{\partial \varepsilon} \right]_{\varepsilon=\varepsilon_F}$, where  $S(\varepsilon)$ is the extremal cyclotron orbit area in $k$-space at Fermi energy $\varepsilon_F$. Following Eqs.~\eqref{onsager} and \eqref{modified_onsager}, we argue that a magnetic field makes the $k$-space cyclotron orbit area $B$-dependent. 
Another physical way to see this is that the magnetic field induces corrections in the band energy of the system, $\varepsilon \to \varepsilon-{\cal M} B + {\cal \kappa} B^2 + ...$, with ${\cal M}$ denoting the magnetic moment and ${\kappa} B^2$ captures the second order correction in the  energy~\cite{Gao_prl14,Gao_prb15,Gao_FP19}. 
%This modification effectively introduces a magnetic field dependence on the cyclotron orbit area $S(\varepsilon-{\cal M} B)$. 
Consequently, the cyclotron mass becomes magnetic field-dependent in some systems. Interestingly, the modified cyclotron area $S(\varepsilon-{\cal M} B + {\cal \kappa} B^2)$ can also be used to determine the correction in the phase factor due to the orbital magnetic moment, see Refs.~\cite{Fuchs2010,gao_PNAS17, Fuch_scipost18}.% for more details. 
%We present two independent pictures of magnetic field-dependent cyclotron mass: semiclassical and quantum.

In the semiclassical picture, we derive the magnetic field dependence of the cyclotron mass by taking the energy derivative of the magnetic field-dependent $k$-space cyclotron area. From Eq.~\eqref{modified_onsager}, we infer that $k$-space area is proportional to the $N(\varepsilon,B)$, the semiclassical IDOS in the presence of $B$. Hence, the field-dependent effective cyclotron mass can be obtained using the formula,
\be \label{cyclotron_mass}
m_{\rm eff}= \frac{\hbar^2}{2 \pi} \left[ \frac{\partial S(\varepsilon,B)}{\partial \varepsilon}\right]_{\varepsilon=\varepsilon_F} \equiv 2 \pi \hbar^2 \left[\frac{\partial N(\varepsilon,B)}{\partial \varepsilon}\right]_{\varepsilon=\varepsilon_F}~.
\ee
%
%Utilizing modified Onsager's relation~\eqref{modified_onsager}, we have 
This yields the field-dependent cyclotron mass to be, %$\left(n + \frac{1}{2}\right) \propto \frac{h}{e B} N_0(\varepsilon) + \frac{h B}{2e} \chi'_0(\varepsilon)$. This implies $N(\varepsilon,B) \propto N_0(\varepsilon) + \frac{1}{2} \chi'_0(\varepsilon) B^2$. Using Eq.~\eqref{cyclotron_mass}, we have
\be
m_{\rm eff}  = 2 \pi \hbar^2  \left[N_0'(\varepsilon_F) + M''_{0}(\varepsilon_F) B + \frac{B^2}{2} \chi''_{0}(\varepsilon_F) + ...\right]~.
%+ \sum_{p\ge3}R''_p(\varepsilon_F) \frac{B^p}{p!} \right].
\ee
The cyclotron mass becomes $B$-dependent as long as any of these quantities, $M''_0(\varepsilon_F)$, $\chi''_0(\varepsilon_F)$, or higher order susceptibility corrections are non-zero. A similar calculation can be done for a three-dimensional system following Eq.~\eqref{Modified_onsager_3D}. In that case, the magnetic-field dependent effective cyclotron mass should be evaluated at extremal values of $k_z=k_{zm}$. We obtain, 
\bea
m_{\rm eff}(k_{zm}) && = \frac{\hbar^2}{ 2 \pi }  \bigg[N_0'(\varepsilon_F,k_{zm}) + M''_{0}(\varepsilon_F,k_{zm}) B \nn \\
&& + \frac{B^2}{2} \chi''_{0}(\varepsilon_F,k_{zm})    + ... %\sum_{p\ge3}R''_p(\varepsilon_F,k_{zm}) \frac{B^p}{p!} 
\bigg]~.
\eea
We note that while the modified Onsager's relation and the resulting nonlinear Landau fan diagram provide insights into the energy derivative of the zero-field magnetic response functions, the variation of $m_{\rm eff}$ with $B$ can serve as a valuable tool to extract the second-order energy derivative of the magnetic response functions.

%The above expression for effective cyclotron mass is very general. It can be used to analyze the field dependency of cyclotron mass once we have the zero-field magnetic response functions obtained either from free energy or the Landau levels. Interestingly, we note that while the modified Onsager's relation and the resulting nonlinear Landau fan diagram provide insights into the energy derivative of the zero-field magnetic response functions, the variation of $m^*$ with $B$ serves as a valuable tool to extract the second-order energy derivative of those response functions.
%However, there is another way to analyze the field dependence of the cyclotron mass---directly using the Landau levels, without delving into the response functions. We discuss this in the following paragraph.

\section{Aperiodic magnetic oscillation: a quantum mechanical perspective} \label{Sec_observables_QM}

Till now, we have predominantly focused on the semiclassical framework. Here, we change gears and focus on the Landau-level framework to understand the nonlinearity in the Landau fan diagram and its consequences.

\subsection{Nonlinear Landau fan diagram}
The nonlinear Landau fan diagram can be constructed based on the modified semiclassical Onsager's quantization condition, as discussed at length in Sec.~\ref{Sec_gen_Onsager}. However, this involves calculating the zero-field response functions, which is often a cumbersome and lengthy task. An independent and alternative way forward is the calculation of the exact LLs. This is also often hindered by the lack of general rules to obtain analytical expressions of the LLs for arbitrary systems. Nevertheless, the LLs can be analytically evaluated in some cases by diagonalizing the Hamiltonian in a magnetic field. Then, we can invert the energy expression to obtain the LL index in terms of energy, magnetic field, and other material-dependent parameters. Assuming a 3D system with an applied magnetic field along the $\hat z$-direction, the LL energies %$\varepsilon_n(k_z,B)$ is a 
are a function of the LL index $n$, $k_z$, and $B$, i.e., $\varepsilon=f(n,k_z,B)$. We can invert the expression $\varepsilon=f(n,k_z,B)$, to obtain the LL index in powers of $B$
%. The $n$ will be a function of $\varepsilon$, $k_z$, $B$ along with other material-specific parameters: 
or $n=g(\varepsilon,k_z,B)$. Expanding this expression in powers of $B$, $n$ can be expressed as 
\bea \label{LL_index_general}
\left(n_{k_z} + \frac{1}{2}\right) \frac{e B}{h} &=& N_0(\varepsilon, k_z) + M'_0(\varepsilon, k_z)B \nn \\
 && + \frac{1}{2}\chi'_0(\varepsilon,k_z) B^2 + \cdots.
\eea%
Theoretically, the Landau fan diagram can be obtained by putting the extremal $k_z$ value in the above equation. Remarkably, we can obtain the physical response functions by performing the $k_z$ sum to add contribution from the whole Fermi pocket,
\be 
\left\{N_0, M'_0, \chi'_0 \right\} = 
\sum_{k_z < k_F} \left\{N_0(\varepsilon,k_z),  M'_0(\varepsilon, k_z), \chi'_0(\varepsilon, k_z)\right\}.
\ee
Here, $k_F$ denotes the Fermi wave-vector.
We discuss the explicit calculation of the physical response functions for the 3D SOC system in Sec.~\ref{Sec_3DSOC}.

\subsection{Aperiodic quantum oscillations in Magneto-conductivity and magnetization}\label{Sec_MR_qm}

In the quantum mechanical framework, the aperiodicity in the magneto-resistivity or the thermodynamic magnetization can be understood by directly calculating it in the presence of quantized Landau levels. Here, we evaluate the longitudinal magneto-conductivity and show that it is proportional to the density of states. Consequently, the aperiodic oscillation in the density of states in $1/B$ reflects in the magneto-conductivity and magnetization.  

In a static electric field, the system is in a non-equilibrium steady state. Using the constant relaxation time approximation, the non-equilibrium distribution function can be expressed as a sum of the equilibrium distribution function ($f^0_n$) and an additional correction ($\delta f_n$). We have $f^0_n \to f^0_n + \delta f_n$ with $\delta f_n = -e \tau {\bm v}_n\cdot {\bm E} (-\partial_{\varepsilon_n} f_n^0)$~. Here, $f_n^0$ specifies the Fermi-Dirac distribution function. Assuming both the electric and magnetic fields are applied along the $z$-direction, the longitudinal current density is given by~\cite{kamal_prr20,Sunit_prb22},
\bea 
%{\bm j} &&= \int [d{\bm k}] (-e) {\bm v} \delta f, \\
j_{z} &=& e^2 \tau  \mathfrak{D} \sum_n \int [dk_z] v_{n,z} {\bm v}_n \cdot {\bm E} (-\partial_\varepsilon f_n^0)~ \nn \\
& =& e^2 \tau \mathfrak{D} E_z \sum_n \int [d k_z] v_{n,z}^2  (-\partial_\varepsilon f_n^0)~.
\eea 
Here, the $[dk_z]\equiv dk_z/{2\pi}$, $\sum_n$ implies the summation over all the Landau levels, and $\mathfrak{D}$ denotes the degeneracy of the Landau levels. The above equation shows that the magneto-conductivity is proportional to DOS convoluted with the square of the magnetic band velocity ($v_n$) in the zero-temperature limit. %Using the Poisson's summation formula [Eq.~\eqref{Poisson_sum}] for sum over $n$, the oscillating part of the longitudinal conductivity ($j_z =\sigma_{zz} E_z$) becomes
% %
% \bea 
% \sigma_{zz} &&= e^2 \tau {\frak D} \sum_p \int [dk_z] \Bigg\{  \cos[2\pi p n_{k_z}]
% \eea 
% %
This is because the DOS is given by the expression 
\be
{\tilde \rho}(\varepsilon,B)=\mathfrak{D} \sum_n \int [dk_z] \delta(\varepsilon -\varepsilon_n(k_z)),
\ee
and $(-\partial_\varepsilon f_n^0)_{T \to 0} = \delta(\varepsilon -\varepsilon_n(k_z))$.
Hence, the aperiodic oscillation in DOS will manifest in magneto-conductivity. Now, the DOS can be calculated from the above expression, or it can also be estimated from the oscillating free energy (${\tilde \Omega}$) by taking its double derivative with energy. We discuss the oscillating part of the free energy and possible aperiodicity in it in Appendix~\ref{app_osc_free}. In particular, we show that the free energy for 3D free electron gas and 3D SOC systems, oscillates periodically and aperiodically, respectively. Hence, the aperiodic oscillation in free energy will directly indicate the aperiodicity in magneto-conductivity, and hence in the magneto-resistivity. Similarly, the oscillation in thermodynamic field-dependent magnetization (${\tilde M}$) can be calculated by taking the derivative with respect to $B$ of the oscillating free energy. %This qualitatively explains the aperiodic oscillation for conductivity and, consequently, magneto-resistivity. 

\subsection{B-dependent effective mass}\label{eff_mass_QM} 
Here, we demonstrate the magnetic field dependence of the effective mass from a quantum mechanical perspective. Let us denote the energy difference between two consecutive LLs with 
\be \label{mass*} 
\hbar \omega_c^*= \eta B^{\nu}/m_{\rm eff}=\varepsilon_{n+1}-\varepsilon_n~, 
\ee
where $m_{\rm eff}$ is the effective cyclotron mass, $\eta$ is the appropriate constant. Here, $\nu$ is the order of $B$ in LLs. For example, for quadratic bands (linear bands) $\nu=1$ ($\nu=1/2$). Note that, in general, $\hbar \omega_c^*$ can contain different order terms in the magnetic field with some material-dependent coefficients. Here, for simplicity, we argue considering only one term with $B^\nu$. Nonetheless, using Eq.~\eqref{mass*} to define the effective cyclotron mass, it becomes $B$-dependent if the energy separation between two consecutive LLs contains terms other than $B^\nu$.

To understand this better, let us consider the example of a 2D electron gas (of mass $m$) in a magnetic field applied perpendicular to the 2D plane. The LL for this system is given by 
$\varepsilon_n=(n+1/2)\hbar \omega_c$. 
Hence, we have 
\begin{subequations}
\bea 
\varepsilon_{n+1}-\varepsilon_{n} &=& \left[(n+3/2)-(n+1/2)\right]\hbar \omega_c~, \\
\implies \hbar \omega_c^* &=& \hbar \omega_c~,~~~~{\rm or}~~~m_{\rm eff} = m~.
\eea
\end{subequations}
Hence, for a 2D electron gas, we have 
$B$-independent effective cyclotron mass. The corresponding Landau fan diagram is linear, and the quantum oscillations are periodic in $1/B$~\cite{Sheonberg}. However, if the separation between consecutive LLs is a nonlinear function of $B^\nu$, then the effective mass becomes a function of $B$ as indicated in Eq.~\eqref{mass*}. We demonstrate the $B$-dependence of $m_{\rm eff}$ explicitly for the spin-orbit coupled 3D electron gas in the next section. 

%i) linear Landau Fan diagram, ii) periodic quantum oscillations 
%The above equation implies $m_{\rm eff}=m$. Therefore, for the 2D free electron gas, the effective cyclotron mass is field-independent and identical to the density of state mass. This is expected since 2D free electron-gas shows periodic magnetic oscillation behavior, where oscillation frequency is magnetic field-independent~\cite{Sheonberg}. Similarly, one can show that pristine graphene and isotropic Weyl semimetals will haave field-independent effective cyclotron mass. Meanwhile, the cyclotron mass for the 3D SOC system is magnetic field-dependent, which we discuss using this method in Sec.~\ref{Sec_3DSOC}. }
%

\section{Nonlinear Landau Fan diagram and aperiodic quantum oscillations in 3D spin-orbit coupled systems} \label{Sec_3DSOC}

\begin{figure}
    \centering
    \includegraphics[width=.9\linewidth]{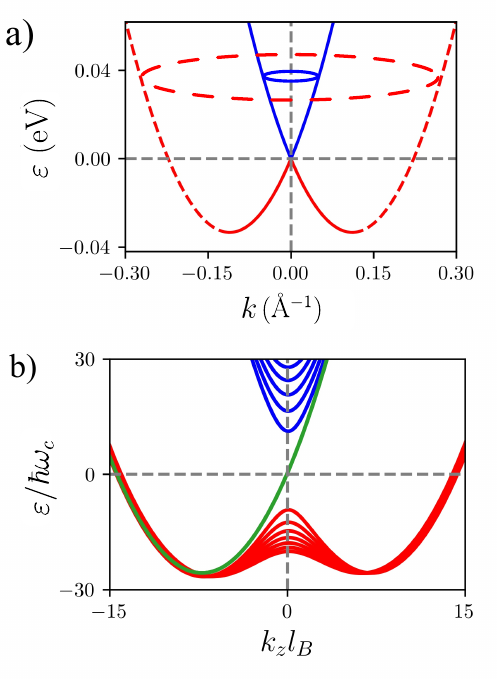}
    \caption{a) The electronic band structure of an isotropic 3D spin-orbit coupled electron gas (for $B=0$). %It has isotropic dispersion. 
    The blue (red) color represents the $\lambda=+1$ ($\lambda=-1$) band. The inner branches are represented by the solid lines (blue for $\lambda=+1$ and red for $\lambda=-1$ band), whereas the outer branches are shown in red dashed lines originating from only $\lambda=-1$ band. b) The Landau level dispersion for the 3D isotropic spin-orbit coupled electron gas in a magnetic field. The blue, red, and green lines represent the conduction band, valence band, and the zeroth Landau level. We have used the following parameters in the above plots: $\alpha=0.8$ eV\AA, $g=2$, and $m=0.067m_e$, where $m_e$ is the electronic mass.}
    \label{fig_3D_disp}
\end{figure}

In this section, we demonstrate a nonlinear Landau fan diagram and aperiodic quantum oscillations in a 3D spin-orbit coupled system. 
%To explore the aperiodic oscillations in a three-dimensional system, we consider a 3D SOC system. Analyzing the aperiodic oscillations involves constructing the modified Onsager's relation, which necessitates the calculation of $k_z$-dependent magnetic response functions. 
%However, in the quantum mechanical framework, 
We directly calculate the exact LLs and invert them to express the LL index $n$ to all orders in the magnetic field. This method proves advantageous when the analytical calculation of the exact LLs is feasible~\cite{Tisserond_IOP17,Fuch_scipost18,rubi_2023unconventional}. 
%In this paper, we opt for this approach. 
The Hamiltonian for the 3D SOC system that we consider is given by~\cite{kang_prb15_transport,samokhin_prb08_effects, law21_comm_phys} 
\be \label{Ham_3D_Rashba}
\mathcal{H} = \frac{\hbar^2 {\bm k}^2}{2m}\sigma_0 + {\alpha} {\bm \sigma}\cdot {\bm k}~.
\ee 
Here, $\bm \sigma = \{\sigma_x, \sigma_y, \sigma_z\}$ denotes the vector composed of the three Pauli matrices representing the real spins, $\sigma_0$ is the $2\times2$ identity matrix, and $\alpha$ is the SOC parameter. The above Hamiltonian has the dispersion relation $\varepsilon_\lambda=\frac{\hbar^2 k^2}{2m}+\lambda k$, where $k=\sqrt{k_x^2+k_y^2+k_z^2}$ and $\lambda=\pm 1$ is the band index. The $\lambda=+1$ ($\lambda=-1$) band is represented by the blue (red) line in Fig.~\ref{fig_3D_disp}a. Both the band touches at $\varepsilon=0$. The $\lambda=+1$ band is monotonic in energy, whereas $\lambda=-1$ band has a nonmonotonic structure with minima at $\varepsilon_{\rm min}=-\varepsilon_\alpha$, with $\varepsilon_\alpha=m\alpha^2/{2\hbar^2}$. The minimum energy lies on the circular contour specified by ${\bm k}^2 = k_\alpha^2$, with $k_\alpha = m \alpha/\hbar^2$. The system has two Fermi contours for both positive and negative energies. For $\varepsilon>0$, the inner Fermi surface originates from the conduction band ($\lambda=+1$), and the outer Fermi surface originates from the valence band ($\lambda=-1$), as shown in Fig.~\ref{fig_3D_disp}a. Also, the carriers in both the Fermi surfaces are electron-like for $\varepsilon>0$ owing to the positive band velocity~\cite{Fuch_scipost18, Sunit_prb23}. However, for $\varepsilon<0$, both the Fermi surfaces originate from the $\lambda=-1$ band. In this case, the inner Fermi surface is hole-like due to the negative band velocity, whereas the outer Fermi surface is electron-like with positive band velocity. In the main text, we work in $\varepsilon>0$ limit, where we can denote the inner (outer) Fermi circles by $\lambda = +1$ ($\lambda = -1$). 
%A similar approach was used in Ref.~\cite{Sunit_prb23} to explain chiral anomaly physics in the 3D SOC system.

%\textcolor{red}{However, we clarify that this notation does not work for the $\varepsilon<0$ limit, where both the Fermi circles originate from the $\lambda=-1$ band. We discuss this explicitly in Appendix~\ref{negative_energy_3DSOC}, see also Ref.~\cite{Sunit_prb23}.}

In the presence of a magnetic field along the $\hat z$ direction, the Hamiltonian \eqref{Ham_3D_Rashba} becomes 
\be \label{Ham_rashba}
\mathcal{H} = \frac{{\bm \Pi}^2}{2 m} \sigma_0
+ \frac{\alpha}{\hbar} {\bm \Pi} \cdot {\bm \sigma} + \Delta_z \sigma_z~.
\ee 
Here, ${\bm \Pi}=\hbar {\bm k} - e {\bm A}$ is the canonical momentum with ${\bm A}$ being the vector potential, and $\Delta_z = g \mu_B B/2$ is the Zeeman coupling energy. We choose the Landau gauge ${\bm A}=(-By,0,0)$ to calculate the LLs. The details of LLs calculation have been outlined in Appendix~\ref{App_LL_calc}. We obtain the LL energies to be 
\begin{widetext}
\begin{subequations}\label{LLs_3DSOC}
\bea 
\varepsilon_{n \neq 0}(k_z) &=& n \hbar \omega_c + \frac{\hbar^2 k_z^2}{2m} + \lambda \sqrt{\frac{2 n \alpha^2}{l_B^2} + \left(\frac{\hbar \omega_c}{2}-{\alpha k_z}  - \frac{g \mu_B B}{2} \right)^2}~, \\
\varepsilon_{n=0}(k_z) &=& \frac{\hbar \omega_c}{2} + \frac{\hbar^2 k_z^2}{2m} - {\alpha k_z} - \frac{g \mu_B B}{2}~.
\eea  
\end{subequations}
Here, $\omega_c = eB/m$ is the cyclotron frequency, and $l_B = \sqrt{\hbar/eB }$ is the magnetic length. The LL spectrums as a function of $k_z$ are shown in Fig.~\ref{fig_3D_disp}b.
Combining the above two equations by introducing a new variable $\Bar{n} = n + \frac{1-\lambda}{2}$~\cite{Fuch_scipost18}, we have  
\be \label{eq:LL_SOC}
\varepsilon_{\bar{n}} = \bar{n} \hbar \omega_c + \frac{\hbar^2 k_z^2}{2m} + \lambda \sqrt{\frac{2 \bar{n} \alpha^2}{l_B^2} + \left(\frac{\hbar \omega_c}{2} -\alpha k_z  - \frac{g \mu_B B}{2} \right)^2}~.
\ee
For $\bar n =0$, there is a single LL included in the $\lambda=+1$ series (with positive root). However, for all $\bar n \geq 1$, there are always two LLs---one with $\lambda=+1$ and the other with $\lambda=-1$. Equation~\eqref{eq:LL_SOC} clearly shows that the LL index is a nonlinear function of $B$. 
%contain all order in $B$ terms, along with the conventional linear $B$ term. %[due to the quadratic $\bm k$ term in Eq.~\eqref{Ham_3D_Rashba}]. 
Now, carefully inverting the above equation, we can write the LL index corresponding to the inner ($\lambda=+1$) and outer ($\lambda=-1$) Fermi surfaces  as 
\begin{subequations}
\bea 
n_{\lambda} &&= \frac{\left(\varepsilon - \frac{\hbar^2 k_z^2}{2m}\right)}{\hbar \omega_c} + \frac{\alpha^2}{(\hbar \omega_c l_B)^2} - \lambda \sqrt{\left(\frac{\alpha}{\hbar \omega_c l_B} \right)^4 + \frac{2 \alpha^2}{(\hbar \omega_c l_B)^2}\frac{\varepsilon}{\hbar \omega_c} + \frac{1}{4} \left( 1 - {\tilde \Delta} \right)^2 -  \frac{\alpha k_z}{ \hbar \omega_c} \left( 1 - {\tilde \Delta} \right) }~, \\
&& \equiv \frac{{C}_0}{B} -\lambda \sqrt{\frac{C_1}{B^2} + C_2 + \frac{C_3}{B}} ~. \label{n_3D_exact}
\eea 
\end{subequations}
\end{widetext}
Here, ${\tilde \Delta}=\frac{g \mu_B B}{ \hbar \omega_c}=g \mu_B m/{e \hbar}$ denotes the ratio between the Zeeman energy and the cyclotron energy. Also, we have defined the coefficients as follows
\begin{subequations}
\be
C_0 = \frac{m}{e \hbar}\left(\varepsilon -\frac{\hbar^2 k_z^2}{2m}  + {2 \varepsilon_\alpha}\right)~, 
\ee %\\
\be 
C_1 = \frac{4 m^2}{e^2 \hbar^2}\left(  \varepsilon_\alpha^2 + \varepsilon \varepsilon_\alpha \right)~,  
\ee 
\be 
C_2 = \frac{1}{4}\left( 1 - {\tilde \Delta} \right)^2 ~, ~~~~{\rm and} 
\ee 
\be 
C_3 = -\frac{m \alpha k_z }{e \hbar} \left( 1 - {\tilde \Delta} \right)~.
\ee
\end{subequations}
Expanding Eq.~\eqref{n_3D_exact} in powers of $B$, we obtain
\bea \label{eq:nb}
n_{\lambda}(\varepsilon, k_z) &=& \frac{C_0 -\lambda  \sqrt{C_1}}{B}  - \frac{\lambda C_3}{2 \sqrt{C_1}} - \frac{1}{2} +\frac{\lambda}{2} \nn \\
& &- \lambda  \frac{(4 C_1 C_2 - C_3^2)}{8 C_1^{3/2}} B + \mathcal{O}(B^2)~.
\eea
%\be \label{eq:nb}
%n_{\lambda} = \frac{C_0 -\lambda  \sqrt{C_1}}{B}  - \frac{\lambda C_3}{2 \sqrt{C_1}} - \frac{1}{2} +\frac{\lambda}{2} - \lambda  \frac{(4 C_1 C_2 - C_3^2)}{8 C_1^{3/2}} B + \mathcal{O}(B^2)~.
%\ee
%
%
%\textcolor{red}{
%It is worth mentioning here that in the above equation, we have introduced the term $- \frac{1}{2} +\frac{\lambda }{2}$ for consistency---
Recall that the $n=0$ LL is considered in the conduction band. For the inner orbit with $\lambda =1$, we have the series $n=0,1,2, \cdots$, whereas for the outer orbit with $\lambda = -1$ we have $n=1,2,3,\cdots$. 
\begin{figure*}
    \centering
    \includegraphics[width=\linewidth]{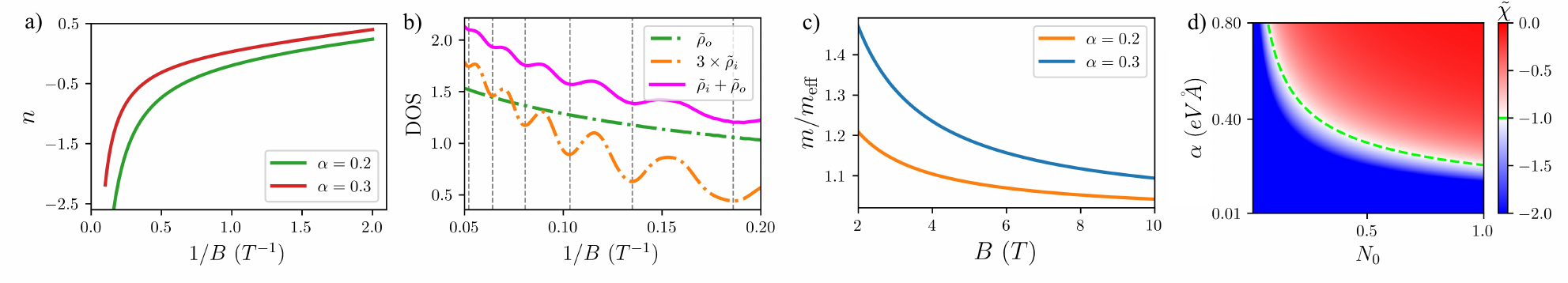}
    \caption{ Landau level-based exact results for a three-dimensional spin-orbit coupled system. 
    a) The nonlinear Landau fan diagram for two different spin-orbit coupling (SOC) strengths [$\alpha$ in units of eV\AA]. b) The density of states  for inner orbit ${\tilde \rho}_i$, and outer orbit ${\tilde \rho}_o$ as a function of $1/B$ for the quantized LLs of Fig.~\ref{fig_3D_disp}b. ${\tilde \rho}_i$ shows clear oscillating behavior with $1/B$ due to distinguishable LLs for the inner Fermi surface ($\lambda=+1$). However,  ${\tilde \rho}_o$ increases monotonically with $B$ without any clear oscillation, as the LLs for the outer Fermi surface ($\lambda = -1$) overlap. More importantly, the magnetic oscillation in ${\tilde \rho}_i$ and in ${\tilde \rho}_i+{\tilde \rho}_o$ are aperiodic in $1/B$, with increasing period as $B$ decreases. 
    c) The decrease of the inverse cyclotron mass ($m_{\rm eff}$) as the magnetic field ($B$) strength grows for two different values of $\alpha$ [$\alpha$ in units of eV\AA]. d) Colorplot of the aperiodicity factor ${\tilde \chi}=\chi'_0/N_0$, showing its variation with SOC strength and carrier density. The green dashed line indicates the  ${\tilde \chi} = -1$ line. %At low carrier density, the aperiodicity factor becomes prominent ($>1$) in the blue region for a given $\alpha$. 
    The blue region with $|{\tilde \chi}| > 1$ is where the aperiodic quantum oscillations and nonlinear landau Fan diagram are likely to be observed. Generally, a smaller Fermi surface with low carrier density is more likely to show a nonlinear Landau fan diagram and aperiodic quantum oscillations due to the $B$-induced changes in the Fermi surface. 
    \label{fig_landau_fan_3D}}
\end{figure*}
%
%
%Owing to the presence of a $B$ linear term in  
We immediately note that Eq.~\eqref{eq:nb} has a form similar to that of generalized Onsager's relation in 3D specified in Eq.~\eqref{Modified_onsager_3D}. Consequently, the $n$ vs. $1/B$ plot for all $k_z$ values deviates from a straight line, producing a nonlinear Landau fan diagram. %We discuss the nonlinear fan diagram for the conduction band while similar logic applies to the valence band. 

Of all the different $k_z$ contributions, the extremal cyclotron orbit contributes the most to the quantum oscillations' aperiodicity and, hence, to the LL fan diagram's nonlinearity. For our 3D SOC model, we can 
see from Fig.~\ref{fig_3D_disp}b that for $k_z=0$, we have $ \partial \varepsilon_n/\partial k_z  =0$ for the inner orbit.
%To plot the Landau fan diagram, in principle, we should find the extremal $k_{z}$ by solving $\left. \partial \varepsilon_n/\partial k_z \right|_{k_{zm}} =0$. However, for LLs of Eq.~\eqref{LLs_3DSOC}, the $k_{zm}$ expression has quite a complicated form and is analytically hard to tackle. 
%Hence, to illustrate the nonlinear Landau fan diagram, we put 
Hence, for $k_z=0$ and the inner Fermi circle ($\lambda=+1$), Eq.~\eqref{eq:nb} reduces to 
%we first find the extremal $k_{z}$ corresponding to the extremal orbit area for which $\left. \frac{\partial S(\varepsilon,k_z)}{\partial k_z} \right|_{k_z=k_{zm}}=0$~\cite{Sheonberg,lifshitz1958theory}. This is equivalent to finding $\frac{\partial}{\partial k_z} (c_0 - \lambda \sqrt{c_1}) =0$ [coefficient of $1/B$ in Eq.~\eqref{eq:nb}]. Similar to the case of 3D free electron gas~\cite{Sheonberg}, in this case as well, we find $k_{zm}=0$. Putting $k_z=k_{zm}=0$ in Eq.~\eqref{eq:nb} yields
%
\bea \label{landau_fan_3D}
n (\varepsilon) & = &\frac{1}{B} \left[\frac{m}{e \hbar}\left(\varepsilon + {2 \varepsilon_\alpha}\right)  - \frac{4 m^2}{e^2 \hbar^2}\left(  \varepsilon_\alpha^2 + \varepsilon \varepsilon_\alpha \right)  \right] \nn \\
& & - {B}\frac{e \hbar}{16 m}  \frac{( 1 - {\tilde \Delta} )^2}{\sqrt{\varepsilon_\alpha^2 + \varepsilon \varepsilon_\alpha}}~.
\eea 
%
% The presence of two Fermi surfaces for Hamiltonian \eqref{Ham_3D_Rashba} signifies the presence of two oscillation frequencies for $\lambda=\pm 1$.
%
We present the corresponding Landau fan diagram in Fig.~\ref{fig_landau_fan_3D}a, highlighting the nonlinear dependence of $n$ with $1/B$. 

Following the discussion in Sec.~\ref{Sec_MR_qm}, the oscillation in physical quantities such as longitudinal resistivity and thermodynamic magnetization arises from the corresponding oscillations in the DOS. Consequently, the aperiodicity in DOS oscillation directly implies that the oscillation of physical quantities will be aperiodic in $1/B$. To explore this, we show the numerically calculated oscillating DOS as a function of $1/B$ for both the inner  ($\lambda=+1$) and outer orbit ($\lambda=-1$) considering  $\varepsilon>0$  in Fig.~\ref{fig_landau_fan_3D}b. The corresponding  LLs are depicted in Fig.~\ref{fig_3D_disp}b. The DOS for outer orbit (${\tilde \rho}_o$) does not show oscillation owing to the almost overlapped LLs for the $\lambda=-1$ branch. Conversely, the DOS for the inner orbit ${\tilde \rho}_i$ exhibits distinct oscillatory behavior with $1/B$. More importantly, we find that the $1/B$ interval between the consecutive troughs or peaks of ${\tilde \rho}_i$ and ${\tilde \rho}_i + {\tilde \rho}_o$ are non-equidistant, signifying aperiodicity of oscillation. We see that the oscillation period (frequency) continuously decreases (increases) with increasing $B$. Additionally, we note that we can obtain the zero-field magnetic responses from Eq.~\eqref{eq:nb}, as we describe below. 

% \subsection{Aperiodic oscillation of DOS}

\subsection{Magnetic response functions}

Having discussed the nonlinear Landau fan diagram,
%and the field-dependent cyclotron mass, 
we now focus on obtaining zero-field magnetic response functions from the LLs. Comparing Eq.~\eqref{eq:nb} with Eq.~\eqref{Modified_onsager_3D}, the $k_z$ dependent response functions are obtained to be 
\begin{subequations}
\bea 
\frac{h}{e} N_{0,\lambda}(k_z) &&= C_0 - \lambda \sqrt{C_1}~, \\
\frac{h}{e} M'_{0,\lambda}(k_z) && = \frac{\lambda}{2} -\lambda \frac{C_3}{2\sqrt{C_1}}~, \\
\frac{h}{e} \chi'_{0,\lambda}(k_z) &&= -\lambda \frac{C_2}{\sqrt{C_1}} + \lambda \frac{C_3^2}{4 C_1^{3/2}}~.
\eea 
\end{subequations}
To extract the zero-field response functions, we need to integrate the corresponding quantities with respect to $k_z$ and sum over both branches. 
%the bands (for $\varepsilon>0$) or the branches (for $\varepsilon<0$). 
The zero-field dispersion relation determines the limits of the $k_z$ integration. Recall that in Eq.~\eqref{eq:nb}, all the coefficients are independent of the magnetic field. We focus on the $\varepsilon>0$ here, while the calculations for $\varepsilon<0$ have been outlined in Appendix~\ref{App_negative_energy_3DSOC}. 
For the inner branch, the limit of integration is given by $k_z \in (- k_i, +k_i)$, where $k_i = 
-k_\alpha  + \sqrt{k_\alpha^2 + {2 m \varepsilon}/{\hbar^2}}$~. 
%to $\frac{m \alpha}{\hbar} - \sqrt{\frac{m^2 \alpha^2}{\hbar^2} + 2 m \varepsilon}$, 
For the outer branch, we have $k_z \in (-k_o, +k_o)$ with $k_o = k_\alpha + \sqrt{k_\alpha^2 + {2 m \varepsilon}/{\hbar^2}}$~. After integrating and summing over both the bands, we obtain the following zero-field magnetic response functions
\begin{subequations}\label{response_3D}
\bea 
\label{Eq:N0}
N_0(\varepsilon) &&= \frac{2 m^3 \alpha^3}{3 \pi^2 \hbar^6} \left( 1  + \frac{\varepsilon}{2 \varepsilon_\alpha}\right)  \sqrt{1 + \frac{\varepsilon}{\varepsilon_\alpha}}~, \\
M'_0(\varepsilon) && = 0~, \\
\chi'_0(\varepsilon) &&=  - \frac{ e^2}{12 \pi^2 m \alpha} \frac{\left( 1 - {\tilde \Delta} \right)^2}{ \left(1 + \frac{\varepsilon}{\varepsilon_\alpha} \right)^{3/2}}~. 
\eea 
\end{subequations}
%
%A few points of interest regarding the above quantities are worth highlighting:
%1)
The calculated IDOS per unit volume in Eq.~\eqref{Eq:N0} corresponds to the analytical result obtained in Ref.~\cite{verma_prb20_dynamical}. Additionally, it reduces to the corresponding expression for the 3D free electron gas in the $\alpha \to 0$ limit. As the Hamiltonian~\eqref{Ham_3D_Rashba} preserves time-reversal symmetry in the absence of a magnetic field, the spontaneous magnetization $M_0(\varepsilon)$ and hence, $M'_0(\varepsilon)$ is zero. For systems with time-reversal symmetry, all the odd-order magnetic response functions vanish~\cite{Glazman_prx18, Fuch_scipost18, Fuch_prb15_orb}. %We show the explicit calculation of $N_0(\varepsilon)$, and $M_0(\varepsilon)$ in Appendix~\ref{negative_energy_3DSOC}.
 The zero-field magnetic susceptibility derivative per unit volume, $\chi'_0(\varepsilon)$, is negative, which introduces the concave-down nonlinearity in the Landau fan diagram. 
 %It would be worth comparing this susceptibility with the expression that can be obtained from linear response theory~\cite{Fukuyama71} or the semiclassical analysis~\cite{Gao_prb15, Gao_FP19}. However, we skip the explicit analytical calculations of $\chi_0 (\varepsilon)$, which is beyond the scope of this work. 
 %\textcolor{cyan}{Nonetheless, we have numerically calculated $\chi(\varepsilon)$ following Ref.~\cite{Gao_prb15}, which shows a similar trend that can be obtained from the Eq.~(\ref{response_3D}c), as shown in Fig.~\ref{?}. We need to show this if we want to have this written. My initial numerical calculation of $\chi$ should give some hints}%Nonetheless, we note that in the limit $\alpha \to 0$, the diamagnetic susceptibility 4

%\textcolor{red}{can we match the susceptibility for 3DEG by taking $\alpha \to 0$? This will be interesting since we are not doing semiclassical calculations for $\chi$. In $\alpha \to 0$, it should reproduce $\chi_0' \propto \sqrt{\varepsilon}$ following Landau diamagnetism of free electron gas--which it does not produce! what went wrong?}

% \subsection{Aperiodic magneto-resistance}

\subsection{$B$-dependent effective cyclotron mass}
We estimate the field-dependent cyclotron mass using the quantum mechanical framework described in Sec.~\ref{eff_mass_QM}. For 3D SOC system, the $\hbar \omega_c^* = \varepsilon_{n+1}(k_z)-\varepsilon_{n}(k_z)$ becomes $k_z$ dependent, following Eq.~\eqref{eq:LL_SOC}. For the inner orbits with $\lambda=+1$, the extremal $k_{zm}=0$. Consequently, at $k_z=0$, we find
\bea 
\hbar \omega_c^* &=& \hbar \omega_c  + \frac{\hbar \omega_c}{2} (1-{\tilde \Delta})^2 \left[ \left\{ 1+\frac{8 (n+1) m^2 \alpha^2 }{e \hbar^3 B (1-{\tilde \Delta})^2} \right\}^{1/2} \right. \nn \\
&& \left. - \left\{ 1+\frac{8 n m^2 \alpha^2 }{e \hbar^3 B (1-{\tilde \Delta})^2} \right\}^{1/2}\right]~.
% \hbar \omega_c^* &=& \hbar \omega_c \left[ 1+ \frac{1}{B}\sqrt{1-\frac{g \mu_B m}{e\hbar}}\frac{\alpha^2 m^2}{2e\hbar^3} + \cdots \right], \\
\eea 
We note that $\hbar \omega_c^*$ contains other $B$-dependent terms than $\hbar \omega_c$. This implies that the cyclotron mass is a function of $B$. Expanding the above equation in powers of $B$, we obtain an approximate expression for $m_{\rm eff}$ to be
\be \label{m_eff_3DSOC}
\frac{1}{m_{\rm eff}} \approx \frac{1}{m} + \frac{1}{B} \frac{2 m \alpha^2 }{e\hbar^3 (1-{\tilde \Delta})}~.
\ee
The above equation implies an increase in the effective mass with the applied magnetic field. A plot of $m/m_{\rm eff}$ vs. $B$ for two different $\alpha$ is shown in Fig.~\ref{fig_landau_fan_3D}c. Similar expressions of field-dependent cyclotron mass were obtained for the two-dimensional Rashba SOC system in Ref.~\cite{rubi_2023unconventional}.

\section{Discussion}\label{Sec_discussion}
In this section, we review recent observations of aperiodic magnetic oscillations and non-linear Landau fan diagrams in 2D and 3D systems. We outline guidelines for observing aperiodic quantum oscillations and discuss other potential mechanisms.
%"Next, we describe some guidelines for systems that show aperiodic quantum oscillations and briefly discuss other mechanisms that can give rise to aperiodic quantum oscillations. 

Prominent examples of the observation of aperiodic magnetic oscillations in 2D systems are Refs.~\cite{Rubi_npj20,rubi_2023unconventional}. These studies explore quantum oscillation on the spin-orbit coupled 2D electron gases at complex oxide interfaces based on SrTiO$_3$ and KTaO$_3$. 
In both studies, the authors observed increased oscillation frequency and cyclotron mass with varying $B$, alongside notable nonlinearity in the Landau fan diagram. 
We observed similar things in our study of SdH oscillations in a 3D spin-orbit coupled electron gas at the KTaO$_3$-based electronic interfaces~\footnote{When the electronic channel width formed at the interface is sufficiently larger than the cyclotron radius of the electron, then we may observe three-dimensional nature of the SdH oscillation. Specifically, we observed SdH oscillation for all three orientations of the applied magnetic field. We elaborately discussed this in our experimental manuscript, which is currently under review.}. In this experimental study, we presented evidence for the magnetic field-induced expansion of the Fermi surface. We explicitly showed that the conventional Lifshitz-Kosevich equation fails to capture the longitudinal magnetoresistivity data, as elaborated in Sec.~\ref{Sec_modified_LK}. 
More importantly, the same samples of KTaO$_3$-based conducting interface exhibit periodic oscillation for higher carrier density samples~\cite{Suvankar_aqt21}. This corroborates our discussion on the condition for smaller density samples or equivalently smaller Fermi surfaces to observe the nonlinear Landau fan diagram in Sec.~\ref{Sec_nl_Fan}.

A compelling question is, what kind of systems will likely show the aperiodic quantum oscillation and nonlinear Landau fan diagram in experiments? We note that having a spinor wavefunction and non-equispaced LLs are not sufficient conditions for systems to display aperiodic quantum oscillation. For example, pristine graphene and ideal Weyl semimetals show periodic oscillation despite having non-equispaced LLs. A nonlinear Landau fan diagram essentially captures the presence of higher-order $B$ terms in the Landau level index $n$ beyond the conventional $1/B$ term. Using this, we identify a few potential scenarios which can lead to these higher-order terms in $n$. Firstly, if the unperturbed system's Hamiltonian contains terms with different orders of crystal momentum, then the Landau levels and LL index will contain different orders of $B$. As an example for 3D systems, in Eq.~\eqref{Ham_3D_Rashba}, we have both $k$-dependent and $k^2$ terms. Another example of a 2D system is gapless graphene with a Zeeman energy term, as discussed in Ref.~\cite{Fuch_scipost18}. The Zeeman energy term creates a gapped Dirac dispersion, and the Hamiltonian has $k^0$ (Zeeman term) and $k$ terms. 
Secondly, cross terms involving mixed products of crystal momentum along different directions (such as $k_x k_y$, for example) make LL
depend on the product of two consecutive LL indices ~\cite{Fuch_scipost18, Chebrolu_23}. This mechanism of nonlinearity in the Landau fan diagram has been demonstrated theoretically in gapped bilayer graphene~\cite{Fuch_scipost18, gao_PNAS17}, and experimentally in twisted double bilayer graphene~\cite{Slot_science23}. %We emphasize that moir\'e materials hosting flat bands will likely show prominent aperiodicity in quantum oscillation. Semiclassically, this can be understood from the fact that the energy derivative of susceptibility $\chi'_0(\varepsilon)$ roughly increases as the square of moir\'e length-scale~\cite{Slot_science23}.    

Our study is valid for the systems in the quantum oscillation regime where many distinguishable Landau levels are occupied. However, 
an extreme quantum limit may exhibit quantum oscillation to be aperiodic in $1/B$, which occurs due to the inversion of the lowest Landau level~\cite{Xie_prb20,xing_2023_rashba}. Our study does not address the magnetic breakdown situation, 
which can also give rise to a nonlinear Landau fan diagram. For example, the nonlinear Landau fan diagram and aperiodic oscillation were recently demonstrated in moir\'e systems~\cite{moon2023nonlinear}. A detailed comparison of these (and possibly other) different mechanisms of generating a non-linear Landau fan diagram is beyond the scope of this paper and deferred to a possible future study. 
%We leave these subtle aspects for future studies. 

%A critical aspect of our study is that it is valid for the systems in the quantum oscillation regime where many distinguishable Landau levels are occupied. Consequently, it does not explain the non-periodic in $1/B$ quantum oscillation beyond the quantum limit, which occurs due to the inversion of the lowest Landau level~\cite{Xie_prb20,xing_2023_rashba}. 
%Furthermore, our study does not encompass the magnetic breakdown scenario. It has been observed in a recent experiment that magnetic breakdown can also give rise to the non-linear Landau fan diagram indicating aperiodic oscillation~\cite{moon2023nonlinear}. We leave the exploration of these subtle aspects of nonlinear Landau fan diagrams for future investigations. 

\section{conclusion} \label{Sec_conclusion}
We developed a detailed theory of the nonlinear Landau fan diagram and aperiodic quantum oscillations in three-dimensional systems. We present both the semiclassical and the quantum mechanical perspectives behind the generalization of Onsager's quantization relation in three-dimensional systems. 
We argue that this is a physical manifestation of the magnetic field-induced expansion/contraction of the Fermi surface of a material, captured by the magnetic susceptibilities induced corrections in the free energy. We showed that the magnetic field-induced Fermi surface changes have a larger impact on systems with a smaller Fermi surface. Consequently, systems with a smaller carrier density are more likely to host the novel phenomena of nonlinear Landau-fan diagrams. 

In addition to producing a nonlinear Landau fan diagram, these magnetic susceptibility corrections induce aperiodic quantum oscillations. Consequently, we argue that the oscillating part of the measured resistivity or the magnetization should be fitted with the generalized Lifshitz-Kosevich formula, that we derive. The modification of the Fermi surface by the magnetic field is also captured by the magnetic field dependence of the effective cyclotron mass extracted from quantum oscillation measurements. We demonstrated all these effects in a 3D spin-orbit coupled system by doing an explicit quantum mechanical calculation. Furthermore, we 
also derive the magnetic response functions of the system from the analytically obtained Landau level spectrum, similar to the 2D case. 
Our work provides novel insights into the fundamental physics of quantum oscillations, with significant implications for the interpretation of experiments.  

\section{Acknowledgements}

S. D. acknowledges the Ministry of Education, Government of India, for financial support through the Prime Minister's Research Fellowship.

\appendix
\begin{widetext}

\section{Aperiodic oscillation in free energy}\label{app_osc_free}
In this Appendix, we sketch the derivation of free energy or the thermodynamic potential using Green's function of the system to illustrate that generalized Onsager's relation leads to aperiodic magnetic oscillation of free energy. Consequently, we argue qualitatively on the aperiodic oscillation in electrical resistance (SdH oscillation) and thermodynamic magnetization (dHvA oscillation). 

The free energy expression can be expressed in terms of Green's function, as discussed by Luttinger and Ward in Ref.~\cite{Luttinger-Ward_1960}. We start with the oscillating part of the free energy, which can be derived from the following expression (see Refs.~\cite{Luttinger-Ward_1960,Adamov_PRB06,kups6418,Pal_prb17,Pal_prbL23} for more details)
% %
% \be 
% \Omega = -k_B T ~{\rm tr}\{ {\rm ln}(- \hat{G}^{-1}) \} - k_B T ~{\rm tr}\{\hat{G} \hat{\Sigma} \} + \Omega'.
% \ee 
%
% \begin{subequations}
\bea \label{free energy_green}
{\tilde \Omega} = -k_B T ~{\rm tr}\left\{{\rm ln}(- \hat{G}^{-1}) \right\} = -\mathfrak{D} k_B T \sum_n \sum_{\omega_m} \sum_{k_z} {\rm ln}(- g_n^{-1}(i\omega_m)).
\eea
% \end{subequations}
%
Here, we have used the fact that the `tr' in Eq.~\eqref{free energy_green} implies the summation over the Landau levels, the fermionic Matsubara frequencies $\omega_m = \pi k_B T (2m +1)$ with $m$ being the integer and other degenerate states within a single Landau level (such as $k_z$ for 3D system or spin). $\mathfrak{D}$ denotes the degeneracy in each Landau level. %The last two terms are introduced to avoid the overcounting of diagrams. However, their oscillatory part cancels each other~\cite{Adamov_PRB06}. Then the magnetic oscillation part is described by 
We have also used ${\rm tr} \{ {\rm ln}({\rm} \hat{G}^{-1}) \} = \sum_i {\rm ln}(g_{i})$, which is true for diagonalizable matrix $\hat{G}$ with eigenvalues $g_{i}$. To express the sum over the LL index $n$ into harmonic sums, we use the Poisson summation formula given by
\be \label{Poisson_sum}
\sum_{n=0}^{\infty} f_n = \int_0^{\infty} dx f(x) + 2 \sum_{p=1}^{\infty} dx f(x) \cos(2 \pi p x).
\ee 
Here, $p$ denotes the harmonic numbers.
Mathematically, the above transformation is only possible when the Fermi energy crosses many distinguishable Landau levels so that the free energy changes smoothly with the magnetic field. Essentially, this happens when the Landau levels are sufficiently broadened due to the impurity scattering or the thermal broadening of the Landau levels.

We consider a three-dimensional system with an external magnetic field applied along the $\hat z$ direction to explicitly discuss the magnetic oscillation in free energy. Consequently, the system's quantized energy levels $\varepsilon_{n}(k_z)$ disperses along the $\hat z$ direction. Now, Green's function is given by
\begin{subequations}
\bea
\hat{G}_n^{-1}(i \omega_m) && = (i \omega_m + \varepsilon_F) - \mathcal{H} - \Sigma , \\
g_n^{-1}(i \omega_m) && = [i \omega_m + \varepsilon_F  - \varepsilon_n(k_z)] + i \frac{\hbar~ {\rm sgn}(\omega_m)}{2 \tau}.~
\eea
\end{subequations}
Here, $\varepsilon_F$ is the Fermi energy. In the above equation, we have ignored the electron-electron interaction effects and considered only the disorder contribution captured by $\Sigma = -i \frac{\hbar~ {\rm sgn}(\omega_m)}{2 \tau}$, $\tau$ being the scattering time~\cite{Adamov_PRB06}. 

Using Poisson's formula in Eq.~\eqref{free energy_green}  yields the oscillating part to be 
\bea 
{\tilde \Omega} &&= - 2 \mathfrak{D} k_B T \sum_{p,\omega_m}  \sum_{k_z} \int dx~ {\rm ln}(- g^{-1}(x, i \omega_m)) \cos(2 \pi p x), \nn \\
&&=  2 \mathfrak{D} k_B T \sum_{p,\omega_m}  \sum_{k_z}  \left[ -\left. {\rm ln} (-g^{-1}(x, i \omega_m)) \frac{\sin(2 \pi x p)}{2 \pi p} \right|_0^{\infty} - \int_0^{\infty} \frac{1}{g^{-1}(x, i \omega_m)} \frac{d}{dx} \left(-g^{-1}(x, i \omega_m) \right) \frac{\sin(2 \pi x p)}{2 \pi p} dx \right]. \label{free_energy_poissons}
\eea 
Note that $\sum_n$ is converted to integration over $x$ in the above expressions. Also, in the last expression, we have done integration by parts with respect to $x$.
The first term of Eq.~\eqref{free_energy_poissons} is a boundary term, which is non-oscillatory. We ignore this term as we focus only on the oscillating part of the free energy. Consequently, the oscillatory term is given by 
\bea 
\tilde{\Omega} && \propto -2 \mathfrak{D} k_B T \sum_{p,\omega_m}  \sum_{k_z} \int_0^{\infty} \frac{1}{g^{-1}(x,k_z,i \omega_m)} \frac{d}{dx} \left(-g^{-1}(x,k_z, i \omega_m) \right) \frac{\sin(2 \pi x p)}{2 \pi p} dx.
\eea 
The above integral can be evaluated using the concept of complex integral, wherein we need to find out the poles of $g^{-1}(x,k_z,i\omega_m)$. The poles are generally complex as we include the disorder scattering and the finite temperature effects. However, we consider a simplified case without considering the disorder scattering and finite-temperature effects. This effectively reduces finding the poles of $g^{-1}(x,k_z,i\omega_m)$ to the solutions of $\varepsilon_x(k_z)=\varepsilon_F$ in terms of $x$. Consequently, we have
\begin{subequations}
\bea 
\tilde{\Omega} && \propto 2 \mathfrak{D} k_B T \sum_{p}  \int {dk_z}~ {\rm Im} \left\{ \int_0^{\infty} \frac{1}{[ \varepsilon_F  - \varepsilon_x(k_z)]} \frac{d}{dx} \left(-\varepsilon_F  + \varepsilon_x(k_z) \right) \frac{e^{i 2 \pi x p}}{2 \pi p} dx  \right\}, \label{free_energy_x_integrand}\\
&& \propto 2 \mathfrak{D} k_B T \sum_{p}  \int {dk_z}~ {\rm Im} \left\{ ~2\pi i ~{\rm Residue}\left[ \frac{1}{[ \varepsilon_F  - \varepsilon_x(k_z)]} \frac{d}{dx} \left(-\varepsilon_F  + \varepsilon_x(k_z) \right) \frac{e^{i 2 \pi x p}}{2 \pi p} \right] \right\}.
\eea 
\end{subequations}
We denote the singularity point/poles of $\varepsilon_x(k_z)=\varepsilon_F$ by $x=x_0$. The $x_0$ is generally a function of $\varepsilon_F$, $k_z$ and $B$. We write the generic version of this as $x_0(\varepsilon_F,k_z, B) = \frac{\mathcal{A}(\varepsilon_F, k_z)}{B} + \mathcal{C}(\varepsilon_F, k_z) + \mathcal{D}(\varepsilon_F,k_z) B + {\cal O}(B^2)$. For example, the LL index for the 3D SOC system contains all orders in $B$ terms; see Eq.~\eqref{n_3D_exact}.
Hence, we obtain
\bea \label{free-enrgy_n_integrated}
\tilde{\Omega} \propto -2 \mathfrak{D} k_B T \sum_{p} ~{\rm Im}\left\{ 2\pi i~ \int {dk_z}  ~  {\frak f}(x_0(\varepsilon_F,k_z,B))\frac{e^{i 2 \pi p  x_0(\varepsilon_F,k_z,B) }}{2 \pi p} \right\}.
\eea 
Here, we have defined ${\frak f}(x_0)= \left. \frac{d}{dx}\varepsilon_x(k_z) \right|_{x=x_0} $. Now, we need to evaluate the $k_z$ integral, for which the model-specific explicit expressions of $x_0(\varepsilon_F,k_z,B)$, and ${\frak f}(x_0)$ is required. We mention that the $k_z$ integration is usually performed with the observation that the above integral contains a rapidly oscillating term; consequently, the largest contribution comes from the extrema of $x_0(\varepsilon_F,k_z,B)$~\cite{Roth66, Sheonberg, lifshitz1958theory, Binghai_23quantum}. Hence, we expand the $x_0(\varepsilon_F, k_z, B)$ in $k_z$ near its extremal points (say, $k_z=k_{zm}$) as follows 
\bea
x_0(\varepsilon_F, k_z, B) &&= \left. x_0(\varepsilon_F, kz, B) \right|_{k_z=k_{zm}} +\left. \frac{1}{2} \frac{\partial^2 x_0(\varepsilon_F, k_z, B)}{\partial {k_z}^2} \right|_{k_z=k_{zm}} (k_z-k_{zm})^2 + \cdots, \nn \\
&& = x_{0m} +\frac{1}{2} x''_{0m} (k_z-k_{zm})^2 + \cdots . \label{tailor_exp_x}
\eea
Here, the $k_{zm}$ is the solution for $\partial_{k_z} x_0(\varepsilon_F, k_z, B)=0$, which is a function of $\varepsilon_F$ and $B$. %In some specific cases, such as the 3D free electron gas $k_{zm}$ is independent of $B$, as discussed in Appendix~\ref{?}.
Replacing the above expression in Eq.~\eqref{free-enrgy_n_integrated}, the oscillating part of the free energy becomes
\be 
\tilde{\Omega} \propto -2 \mathfrak{D} k_B T \sum_{p} ~{\rm Im} \left\{ 2\pi i~ \int {dk_z}  ~{\frak f}\left(x_{0m} \pm x''_{0m} (k_z-k_{zm})^2/2 \right) \frac{{\rm exp}\left[i 2 \pi p  \left(x_{0m} \pm x''_{0m} (k_z-k_{zm})^2/2 \right) \right]}{2 \pi p} \right \}.
\ee 
We note that the $x_0(k_z, \varepsilon_F, B)$ can contain terms to all orders in $B$. So, we expect that $x_{0m} = x_0(k_z=k_{zm}, \varepsilon_F, B)$ will also contain terms to all orders in $B$, which will again have the general form $\frac{{\cal A}(\varepsilon_F)}{B} + {\cal C}(\varepsilon_F) + {\cal D}(\varepsilon_F) B + {\cal O}(B^2)$. With this, the free energy expression can be rewritten as
\be \label{free_energy_final}
\tilde{\Omega} \propto - 2 \mathfrak{D} k_B T \sum_{p} ~{\rm Im} \left\{ 2\pi i e^{i 2\pi p \left(\frac{{\cal A}(\varepsilon_F)}{B} + {\cal C}(\varepsilon_F) + {\cal D}(\varepsilon_F) B\right)} \int {dk_z}~  {\frak f} \left(x_{0m} \pm x''_{0m} (k_z-k_{zm})^2/2 \right) \frac{{\rm exp}\left[\pm i 2 \pi p 
 x''_{0m} (k_z-k_{zm})^2/2 \right]}{2 \pi p} \right \},
\ee 
where we have explicitly used $x_{0m} \approx \frac{{\cal A}(\varepsilon_F)}{B} + {\cal C}(\varepsilon_F) + {\cal D}(\varepsilon_F) B$. Finally, we need to evaluate the $k_z$ integral to obtain the expression for $\tilde \Omega$ for any specific system. {Nonetheless, the most important aspect of the above equation is the presence of the term $e^{i 2 \pi p x_{0m}}$, which indicates possible aperiodic oscillation if the coefficient of $B$ in $x_{0m}$, i.e., ${\cal D}(\varepsilon_F) $ is non-zero.} Below, we show that free energy for 3D free electron gas oscillate periodically with $1/B$, owing to the ${\cal D}(\varepsilon_F)=0 $. However, free energy for the 3D SOC system will show aperiodicity in $1/B$, as our heuristic calculation indicates, with exact calculation being analytically intractable.

{\bf 3D free electron gas:---}
For 3D free electron gas with a magnetic field along the $\hat z$-direction, we have the LLs $\varepsilon_n(k_z)=(n+1/2)\hbar \omega_c +\hbar^2 k_z^2/2m$. Consequently, we have $x_0(\varepsilon_F, k_z, B)=\frac{1}{\hbar \omega_c} \left(\varepsilon_F - \frac{\hbar^2 k_z^2}{2m} \right) - \frac{1}{2}$. The extrema of $x_0$, obtained from $\partial_{k_z} x_0(\varepsilon_F, k_z, B)=0$ is found to be $k_{zm}=0$. Using Eq.~\eqref{tailor_exp_x}, the $x_0(\varepsilon_F, k_z, B)$ can be expressed as 
\bea
x_0(\varepsilon_F, k_z, B)  = x_{0m} + \frac{1}{2}x''_{0m} k_z^2 + \cdots ,
\eea
with $x_{0m} =\frac{1}{B} \frac{m \varepsilon_F}{e \hbar}-\frac{1}{2}$. Also, we have ${\frak f}(x_0) =\frac{d}{dx}\varepsilon_x(k_z)= \hbar \omega_c$.
Using these for free electron gas, the oscillating part of the free energy becomes
\bea 
\tilde{\Omega} &&\approx 2 \mathfrak{D} k_B T \hbar \omega_c \sum_{p} ~{\rm Im} \left\{ 2\pi i~ \int dk_z  ~ \frac{{\rm exp} \left[i 2 \pi p  (x_{0m} \pm x''_{0m} k_z^2/2) \right]}{2 \pi p} \right\}, \nn \\
&& \approx 2 \mathfrak{D} k_B T \hbar \omega_c \sum_{p} \frac{1}{ p} \int dk_z \left[ \cos( 2 \pi p x_{0m}) \cos(2 \pi p x''_{0m} k_z^2/2) \mp \sin( 2 \pi p x_{0m}) \sin(2 \pi p x''_{0m} k_z^2/2) \right]
\eea
Here, the $\pm$ is the sign of $x''_{0m}$, where the $+$ ($-$) sign represents that $x_{0m}$ is a minimum (maximum).
The above integral can easily be evaluated by introducing a change of variable $u^2=x''_{0m} k_z^2/2$, where we use the $k_z$ limits to be $k_z \in (-\infty, \infty)$. The zero-field dispersion relation should be used to determine the $k_z$ limit, as in Sec.~\ref{Sec_3DSOC}. However, the above limits are a good approximation when $k_z$ is measured from the extremal point of $x_0(\varepsilon, k_z, B)$~\cite{Sheonberg, Roth66}. Finally, using the identity $\int_0^{\infty} \cos(\pi u^2/2)=\int_0^{\infty} \sin(\pi u^2/2) = 1/2$, we have
\be 
\tilde{\Omega} \approx 2 \mathfrak{D} k_B T \hbar \omega_c \sum_{p} \frac{\sqrt{\pi/2}}{( 2 \pi p x''_{0m})^{1/2}}  ~ \frac{ \cos(2\pi p (\frac{1}{B} \frac{m \varepsilon_F}{e \hbar}-\frac{1}{2}) \pm \frac{\pi}{4})}{ p}.
\ee 
The above equation shows that the free energy oscillates periodically with $1/B$. Consequently, the density of states and thermodynamic magnetization, evaluated as ${\tilde \rho} =- \frac{\partial^2 {\tilde \Omega}}{\partial \varepsilon_F^2}$ and ${\tilde M} = \frac{\partial {\tilde \Omega}}{\partial B}$, also oscillates periodically. 

{\bf 3D SOC model:---}
Like free electron gas, we start with the solution of $\varepsilon_n(k_x)=\varepsilon_F$. However, the solution of $\varepsilon_x(k_z) = \varepsilon_F$ in terms of $x$ admits infinite series expansion in $B$ for the Hamiltonian~\eqref{Ham_3D_Rashba}. In Eq.~\eqref{eq:nb}, we have the complete expression up to linear order in $B$ term. Consequently, we write the generic version of this as $x_0(\varepsilon_F,k_z, B) = \frac{{\cal A}(\varepsilon_F, k_z)}{B} + {\cal C}(\varepsilon_F, k_z) + {\cal D}(\varepsilon_F,k_z) B + {\cal O}(B^2)$. %Hence, we will work with this approximated $x_0$ for the integration over $x$. 
Also, we have
\be
{\frak f}(x) = \frac{d}{dx}\varepsilon_x(k_z) =  \hbar \omega_c + \lambda \frac{2\alpha^2/l_B^2}{\sqrt{x \frac{2 \alpha^2}{l_B^2} + \left(-{\alpha k_z} + \frac{\hbar \omega_c}{2} - \frac{g \mu_B B}{2} \right)^2}}.
\ee 
Unlike the free electron case, we see that the ${\frak f}(x)$ depends on the $k_z$. The complicated form of ${\frak f}(x)$ obstructs us from using the Gaussian integral technique, similar to the case of free electron gas. %Moreover, the extremal $k_z=k_{zm}$ is not zero, but rather some complicated function of $\varepsilon_F$ and $B$. 
Nonetheless, the oscillating part of the free energy becomes
\be 
\tilde{\Omega} \approx 2 \mathfrak{D} k_B T \sum_{p} ~{\rm Im} \left\{ 2\pi i~ \int dk_z  ~{\frak f}\left(x_{0m} \pm x''_{0m} (k_z-k_{zm})^2/2 \right) \frac{{\rm exp}\left[i 2 \pi p  \left(x_{0m} \pm x''_{0m} (k_z-k_{zm})^2 /2 \right) \right]}{2 \pi p} \right \}.
\ee 
We note that $x_0(k_z, \varepsilon_F, B)$ already contains the terms other than $1/B$. So, the $x_{0m} \approx x_0(k_z=k_{zm}, \varepsilon_F, B)$ will also contain terms with $B$, which will have general form $\frac{{\cal A}(\varepsilon_F)}{B} + {\cal C}(\varepsilon_F) + {\cal D}(\varepsilon_F) B$. This can already be seen from Eq.~\eqref{landau_fan_3D}, which shows that the LL index contains linear order terms in $B$ for the inner orbit. With this, the free energy expression can be written as
\be \label{free_energy_3DSOC_final}
\tilde{\Omega} \approx 2 \mathfrak{D} k_B T \sum_{p} ~{\rm Im} \left\{ 2\pi i e^{i 2\pi p \left(\frac{{\cal A}(\varepsilon_F)}{B} + {\cal C}(\varepsilon_F) + {\cal D}(\varepsilon_F) B\right)} \int dk_z  {\frak f}\left(x_{0m} \pm x''_{0m} (k_z-k_{zm})^2 /2 \right) \frac{{\rm exp}\left[\pm i 2 \pi p 
 x''_{0m} (k_z-k_{zm})^2/2 \right]}{2 \pi p} \right \},
\ee 
where we have explicitly used $x_{0m} \approx \frac{{\cal A}(\varepsilon_F)}{B} + {\cal C}(\varepsilon_F) + {\cal D}(\varepsilon_F) B$, and the $k_z$-integral can be evaluated numerically. The presence of the non-zero coefficient of $B$ term makes the oscillation in the free energy aperiodic, which will be reflected in the magnetization and density of states. This establishes that the 3D SOC system will show aperiodic oscillation in physical quantities.

\section{Details of the LLs calculation for 3D SOC system} \label{App_LL_calc}
The magnetic field along the $\hat z$-direction is represented by the vector potential ${\bm A}=(-By, 0, 0)$. Consequently, the translation symmetry remains invariant along the $x$- and $z$-direction. Hence, we look for the solution of the form ${\cal H} \Psi = \varepsilon_n \Psi$, where $\Psi= \psi(y) e^{i (p_x x + p_z z)/\hbar}$. With such a plane wave basis, the Hamiltonian is modified as 
\bea
{\cal H} && = 
\begin{pmatrix}
\frac{(  {\bm p} -  e {\bm A}) ^2}{2 m} +  \frac{\alpha}{\hbar} p_z + \Delta_z   & \frac{\alpha}{\hbar} (p_x - eB y - i {\hat p}_y)
\\[2ex]
\alpha (p_x - eB y + i {p}_y) & \frac{( {\bm p} - e {\bm A})^2}{2 m}- \frac{\alpha}{\hbar} p_z - \Delta_z
\end{pmatrix}, \nn \\
&&=\begin{pmatrix}
\frac{\hbar^2}{m l_B^2}(\hat{a}^\dagger \hat{a}+ \frac{1}{2}) + \frac{p_z^2}{2m} + \frac{\alpha}{\hbar} p_z + \Delta_z   & \frac{\sqrt{2} \alpha}{l_B} \hat{a}
\\[2ex]
\frac{\sqrt{2} \alpha}{l_B} \hat{a}^\dagger & \frac{\hbar^2}{m l_B^2}(\hat{a}^\dagger \hat{a} + \frac{1}{2}) + \frac{p_z^2}{2m} - \frac{\alpha}{\hbar} p_z - \Delta_z
\end{pmatrix}. 
\eea
Here, we have introduced a new variable ${\tilde y} = (y/l_B - p_x l_B/\hbar)$, and subsequently, the creation and annihilation operators ${\hat a}^\dagger =  ({\tilde y} - \partial_{\tilde y})/\sqrt{2} $ and ${\hat a} = ({\tilde y} + \partial_{\tilde y})/\sqrt{2} $, satisfying the commutation relation $[{\hat a}, {\hat a}^\dagger]=1$. Also, we have used the relation $\left[(p_x-eBy)^2 + p_y^2 \right]= \frac{2 \hbar^2}{l_B^2} (\hat{a}^\dagger \hat{a}+ \frac{1}{2})$. Now, one can obtain the LLs for the above Hamiltonian by solving the eigenvalue equation, using the spinor i) $\psi(y) = [0 ~~ {\cal N}_0 \psi_{0}]^T$ when $n=0$, and ii) $\psi(y) = [{\cal N}_1  \psi_{n-1}~~ {\cal N}_2 \psi_{n}]^T$ when $n \neq 0$. Here, $\psi_n$ are the usual harmonic oscillator wave functions, and ${\cal N}_{0,1,2}$ are the normalization constants. Utilizing the relations: $a^\dagger \psi_n = \sqrt{n+1} \psi_{n+1}$, $a \psi_n = \sqrt{n} \psi_{n-1}$, and $a^\dagger a \psi_n = {n} \psi_{n-1}$, we can easily obtain the LLs of Eq.~\eqref{LLs_3DSOC}.

\section{The LL index for 3D SOC system in $\varepsilon<0$ regime}\label{App_negative_energy_3DSOC}
% {\bf $\varepsilon_F<0$ case:---} 
For $\varepsilon<0$, there is only one band (valence) as shown in Fig.~\ref{fig_3D_disp}a. However, in this case, both electron-like and hole-like carriers exist. The Fermi surface contour is made of two circles, where the inner circles (we will represent it by $b=-1$) represent the hole-like carriers. In contrast, the outer circle ($b=+1$) represents the electron-like carriers. Here, electron-like (hole-like) carriers are equivalent to having positive (negative) band velocity (${\bm v}_\lambda={\bm \nabla}_{\bm p} \varepsilon_\lambda$). As a result, among the LLs, some are electron-like (corresponding to the outer branch of the Fermi circle), and some are hole-like (corresponding to the inner branch of the Fermi circle)~\cite{Fuch_scipost18}. In the Landau-level picture, the LLs whose energy increases (decreases) with LL index $n$ are called electron-like (hole-like). The expression for both types of LL index can be obtained from Eq.~\eqref{n_3D_exact} by replacing $\lambda$ with $-b$. Then, we have
\be \label{n_3d_hole}
n_b(\varepsilon,k_z) = \frac{C_0 + b \sqrt{C_1}}{B}  + \frac{b C_3}{2 \sqrt{C_1}} - 1 + b \frac{(4 C_1 C_2 - C_3^2)}{8 C_1^{3/2}} B + \mathcal{O}(B^2)~.
\ee
%  
% We can understand this because hole-like carriers correspond to lower energy (for $k_F<k<k_\alpha$) and hence the lower LL index. In contrast, the electron-like LLs correspond to higher energy ($k_\alpha<k<k_F$) and hence the higher LL index. This can be visualized from Fig.~4c of Ref.~\cite{Fuch_scipost18}. 
For electron-like LLs, we can directly compare the above equation with Eq.~\eqref{modified_onsager} and obtain the $k_z$-dependent zero-field magnetic response function as follows:
\begin{subequations}  \label{kz_res_o}
\bea 
\frac{h}{e} N_{0,b=+1}(k_z) && = C_0 + \sqrt{C_1}, \\
\frac{h}{e} M'_{0, b=+1}(k_z) && = -\frac{1}{2} + \frac{C_3}{2\sqrt{C_1}}, \\
\frac{h}{e} \chi'_{0,b=+1}(k_z) && =  \frac{C_2}{\sqrt{C_1}} - \frac{C_3^2}{4 C_1^{3/2}}.
\eea 
\end{subequations}
%
% The integration limit in this case is given by $k_z \in (-k_v, +k_v)$ with $k_v = \frac{m \alpha}{\hbar} + \sqrt{\frac{m^2 \alpha^2}{\hbar^2} + 2 m \varepsilon}$.
However, for hole-like LLs ($b=-1$), we need to compare the above equation with~\cite{Fuch_scipost18}
\bea 
\left(n+\frac{1}{2} \right)\frac{eB}{h} = {N_{\rm tot}-N(\varepsilon,k_z,B)}= {N_{\rm tot}-N_0(\varepsilon,k_z)} - B M'_0(\varepsilon,k_z) - \frac{B^2}{2} \chi'_0(\varepsilon,k_z) + \cdots.
\eea 
We mention that for three-dimensional systems, $N_{\rm tot}$ is the $k_z$ resolved total number of electrons when the band is fully occupied. While working with the low-energy model, $N_{\rm tot}$ is usually determined from self-consistency~\cite{Fuch_scipost18}.  Actually, $N(\varepsilon,k_z, B) = \int d\varepsilon \rho(\varepsilon,k_z,B) f_0$ is being replaced by $N_{\rm tot}-N(\varepsilon,k_z, B) = \int d\varepsilon \rho(\varepsilon,k_z,B) (1-f_0)$, where $\rho(\varepsilon,k_z, B)$ is the $k_z$ resolved DOS, and $f_0$ is the equilibrium Fermi function.

Now, comparing the above equation with Eq.~\eqref{n_3d_hole}, the $k_z$ dependent response functions are obtained to be
\begin{subequations} \label{kz_res_i}
\bea 
\frac{h}{e} \left[N_{\rm tot}-N_{0,b=-1}(k_z) \right] && = C_0 - \sqrt{C_1}, \\
\frac{h}{e} M'_{0, b=-1}(k_z) && = \frac{1}{2} + \frac{C_3}{2\sqrt{C_1}}, \\
\frac{h}{e} \chi'_{0,b=-1}(k_z) && =  \frac{C_2}{\sqrt{C_1}} - \frac{C_3^2}{4 C_1^{3/2}}.
\eea 
\end{subequations}
To evaluate the physical response functions, the $k_z$ integration limit are given by $k_z \in \{-k_{i},+k_{i}\}$ for the hole-like ($b=-1)$ branch, where $k_{i} = -k_\alpha + \sqrt{k_\alpha^2 + 2 m \varepsilon/\hbar^2}$. For the electron-like ($b=+1$) branches, the limit is given by $k_z \in \{-k_{o}, k_{o} \}$, where $k_{o} = k_\alpha + \sqrt{k_\alpha^2 + 2 m \varepsilon/\hbar^2}$. 
Upon integrating quantities in Eqs.~(\ref{kz_res_o} and \ref{kz_res_i}) and adding the results, we can obtain similar expressions for the magnetic response functions. We find that Eq.~\eqref{response_3D} still gives zero-field response functions. This establishes the consistency of our calculations.

\end{widetext}

\bibliography{ref}

\end{document}